\documentclass[epj]{svjour}
\usepackage{graphics}
\usepackage[T1]{fontenc}
\usepackage[latin1]{inputenc}
\usepackage{amsmath}
\usepackage{graphicx}
\usepackage{amssymb}

\usepackage{graphicx}
\usepackage{epsfig}
\usepackage{multirow}
\usepackage{dcolumn}
\usepackage{bm}
\usepackage{subfigure}

\def\<{\langle}
\def\>{\rangle}

\begin{document}
\title{Linear and nonlinear transport across carbon nanotube quantum dots}

\author{Leonhard Mayrhofer \and Milena Grifoni% etc
}   

\institute{Theoretische Physik, Universit\"{a}t Regensburg, 93040 Regensburg, Germany}
\date{Received: date / Revised version: date}
\abstract{
We present a low energy-theory for non-linear transport in finite-size
interacting single-wall carbon nanotubes. It is based on a microscopic
model for the interacting $p_{z}$ electrons and successive bosonization.
We consider weak coupling to the leads and derive equations of motion
for the reduced density matrix. We focus on the case of large-diameter nanotubes where exchange effects can be neglected. In this situation the energy spectrum is highly degenerate. Due to the multiple degeneracy, diagonal as well as off-diagonal (coherences)
elements of the density matrix contribute to the nonlinear transport.
At low bias, a four-electron periodicity with a characteristic ratio
between adjacent peaks is predicted. Our results are in quantitative
agreement with recent experiments. 
 % end of PACS codes
} %end of abstract
\maketitle

\section{Introduction\label{sec:Introduction}}

Single-walled carbon nanotubes (SWNTs) are graphene sheets rolled
up to seamless cylinders, which possess spectacular mechanical and
electrical properties \cite{Saito1998,Loiseau2006}. In particular, as suggested
in the seminal works \cite{Egger1997,Kane1997,Yoshioka1999}, due
to the peculiar one-dimensional (1D) character of their electronic bands,
metallic SWNTs are expected to exhibit Luttinger liquid behavior
at low energies, reflected in power-law dependence of various quantities
and spin-charge separation. Later experimental observations have provided
a confirmation of the theory \cite{Bockrath1999,Postma2001}. 

\begin{figure}
\includegraphics[%
  width=1.0\columnwidth]{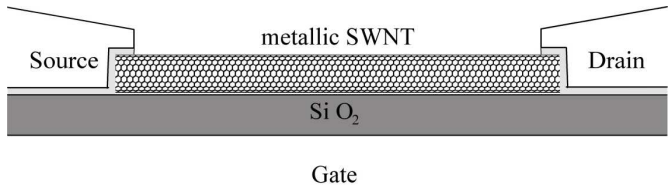}

\caption{\label{cap:SWNT-quantum-dot}SWNT quantum dot setup. A metallic SWNT
is coupled via tunnelling junctions to a source and a drain contact.
The electrochemical potential in the dot is adjusted by a gate voltage.}
\end{figure}

In order to study the internal electronic properties, including the
effects of electron - electron correlations, a quantum dot
setup, where a SWNT is coupled weakly to a source and drain contact
as well as capacitatively to a gate electrode (cf. Fig. \ref{cap:SWNT-quantum-dot}),
is a very well suited device. In fact, the current as a function of the
applied bias voltage $V_{b}$ and gate voltage $V_{g}$ depends on
the energy spectrum but also on the actual form of the eigenstates
of the considered system, which for example determines whether some transitions are allowed or forbidden. At low enough temperatures and bias voltage, adding new particles --- and therefore transport across
the quantum dot --- is hindered by the Coulomb repulsion \cite{Tans1997}.
Eventually the energy cost for placing an electron in a given energy
level adds to the effect of Coulomb blockade. 
In metallic SWNTs two bands cross at the Fermi energy. Together with the spin degree of freedom this leads to the formation of electron shells, each accommodating up to four electrons.

%By taking into account that in metallic SWNTs two bands cross at the Fermi energy, and including
%the spin degrees of freedom, this results in characteristic even-odd 
As a result, a characteristic even-odd
\cite{Cobden2002} or fourfold \cite{Liang2002,Sapmaz2005,Moriyama2005}
periodicity of the Coulomb diamond size as a function of the gate
voltage is found. While the Coulomb blockade can be explained merely by the
ground state properties of a SWNT, the determination of the current
at higher bias voltages requires the inclusion of transitions of the
system to electronic excitations. For an interacting 1D system the
occurrence of spin-charge separation is expected, i.e. the collective
excitations can be divided into independent modes with different spectra.
In the case of metallic SWNTs one finds three so called neutral modes \cite{Kane1997},
whose energies coincide with the corresponding ones of the noninteracting
system. Of the three modes two can be identified as collective spin
excitations and one as collective charge excitations. Additionally,
there is one mode of charge excitations with energies increased compared
to the neutral modes due to the repulsive electron - electron interaction.
In \cite{Sapmaz2005,Moriyama2005} not only Coulomb diamonds but also low lying
excitation lines have been resolved in $I$ - $V_{b}$ - $V_{g}$
measurements. Since the associated energies are not large enough in
order to identify some of the observed lines with transitions to excitations
of the interaction-dependent mode, the position of all the measured
lines should be explainable by invoking the excitation spectrum of the system with neutral modes only.

Additionally in \cite{Moriyama2005} the effect of an applied magnetic field on the transport properties of a SWNT quantum dot was examined. Clear evidence for an exchange splitting of the ground states with two electrons in the highest occupied shell was found. 
Exchange effects were also inferred from the excitation lines of the spectrum of samples A and B of \cite{Sapmaz2005}, in agreement with predictions of the mean-field theory presented in \cite{Oreg2000}.
%Such a behavior has been predicted within a mean-field theory \cite{Oreg2000}. 
On the other hand, the measured data of sample C in \cite{Sapmaz2005} do not show any sign of exchange effects, unless an unreasonably high exchange energy is assumed.
The magnitude of the exchange effects depends on the spatial extension of the electron wave functions and scales like $1/N_{\mathrm{atoms}}$, where $N_{\mathrm{atoms}}$ is the number of carbon atoms in the particular SWNT \cite{Leo2007}. Therefore, e.g. the ratio between the level spacing $\varepsilon_0$ of a noninteracting SWNT, scaling like $1/L$, and the exchange related energies is roughly proportional to $D$, where $L$ and $D$ are the nanotube length and diameter respectively. On that score it is interesting to note that the experiments in \cite{Moriyama2005}, where the exchange splitting could be seen, was conducted with a SWNT of $L=300$ nm and a comparably small diameter of $0.8$ nm. Similarly for samples A and B of \cite{Sapmaz2005} the reported lengths and diameters were $L_{A}=180$ nm and $D_{A}=1.1$ nm and $L_{B}=500$ nm and $D_{B}=1.3$ nm. In contrast, for the SWNT of sample C in \cite{Sapmaz2005}, showing no measurable exchange effects, $L_{C}=750$ nm and $D_{C}=2.7$ nm were determined. For comparison, in the case of armchair tubes the diameters $0.8$ nm and $2.7$ nm would correspond to tubes with the wrapping indices $(6,6)$ and $(20,20)$, respectively. In the following we concentrate on large enough nanotubes and disregard any exchange effects. A detailed discussion of the processes leading to exchange splittings and their interplay with the bosonic excitations is postponed to a forthcoming article \cite{Leo2007}.
In the absence of exchange, a large degeneracy of the energy spectrum is expected \cite{Leo2006}, which in turn can be seen in a peculiar four-electron periodicity of the stability diagrams (three equal in size Coulomb diamonds followed by a larger one) and of the Coulomb oscillation traces as discussed below.

%So far, only mean-field theories have been proposed to account
%for the excitation spectrum \cite{Ke2003,Oreg2000,Bellucci2005}. In
%particular, a good agreement of the measurements with the mean field
%predictions of \cite{Oreg2000} was stated in \cite{Sapmaz2005}.
%However, a mean field description may not be justified for 1D systems.
%A comparison between the outcomes of our bosonization approach and
%of mean field theory is presented in this work. We find that the low
%bias spectrum can be described equally well within the two theories.
%However, qualitative differences occur for finite bias voltages.

%

So far excitation lines of SWNT quantum dots were only addressed in \cite{Oreg2000,Ke2003,Bellucci2005} in a meanfield approach.
In the present work we do not only calculate the expected excitation lines
but also give a quantitative calculation of the
nonlinear current across a metallic SWNT quantum dot as a function
of the gate and bias voltage beyond meanfield. As long as only the ground state energies are concerned, the meanfield treatment yields the right result but fails for a complete description of the excitation spectrum. Some results were already presented in abridged form in \cite{Leo2006}. 
We focus on the low energy regime, which means that we consider the situation where only the gapless
subbands of the SWNT are relevant. Since those bands exhibit to a
good approximation a linear dispersion relation, the Tomonaga-Luttinger
theory can be applied. For a typical SWNT this means a range of roughly $1$ eV around the Fermi
energy, which is still quite large if for example compared to the level spacing
of $1.7$ meV for a noninteracting SWNT of 1 $\mu$m length.

Starting from a microscopic description of noninteracting SWNTs with
open boundary conditions we include the electron - electron interactions.
The resulting Hamiltonian can be diagonalized using bosonization techniques 
\cite{Egger1997}, obtaining the spectrum and eigenstates of the isolated
metallic finite size SWNT. We apply the so called constructive bosonization
procedure \cite{Haldane1981} (for a detailed review we refer to \cite{Delft1998}),
coping well with the discrete energy spectrum and especially with
finite electron numbers.

The electron dynamics of the quantum dot is obtained by solving the
equation of motion for the reduced density matrix (RDM) of the SWNT.
The metal leads are described as equilibrated Fermi gases. Due to many-fold degenerate eigenstates
of the SWNT Hamiltonian we find that offdiagonal elements (\textit{coherences})
of the RDM can not be generally ignored. Apart from the degeneracy of
the states, this is also a consequence of the interactions as we will
show. The importance of taking into account coherences for an interacting
system weakly coupled to unpolarized leads was also recently discussed
in \cite{Braig 2005}, where transport through a metal grain is treated.
The rates entering the master equation depend on the involved energies
and on the matrix elements of the electron operators in the SWNT eigenstate
basis. Noteworthy is a variation of the tunnelling amplitudes along
the nanotube axis, also if only ground states are involved in transport,
depending on the complete energy spectrum of the collective excitations.

The outline of this article is the following. In Section \ref{sec:SWNT-quantum-dots}
we derive the master equation describing the dynamics of a generic
quantum dot in second order of the tunnelling, gaining an expression
for the current. In order to specify the explicit form of the master equation we derive
the low energy Hamiltonian of metallic SWNTs in Section \ref{sub:Low-energy-description},
and use the bosonization technique in order to diagonalize the SWNT
Hamiltonian. The actual calculations for the current in the low and
high bias regime are performed in Section \ref{sub:Current-trough-a}.

\section{Quantum dots\label{sec:SWNT-quantum-dots}}

In this section we show how the stationary current through the system
described by the Hamiltonian (\ref{eq:H_tot}), see below, can be determined as
a function of the electrochemical potentials in the leads and in the
dot by calculating the dynamics of the RDM of the SWNT. Transport
through the SWNT quantum dot can occur if the electrochemical potentials
of source and drain are adjusted to different values by the voltages
$V_{s}$ and $V_{d}.$ Since current can only flow if one transport
direction is favored, the current-carrying system will have to be
treated out of equilibrium beyond the linear response regime. Hence, 
the state of the quantum dot will in general be described by a nonequilibrium
density matrix. For completeness and clarity, we show the determination
of the corresponding equation of motion in some detail mainly following 
\cite{Blum1996}. The outcomes of this and of the following section
will be used in Section \ref{sub:Current-trough-a} to obtain the
$I$-$V_{b}$-$V_{g}$ characteristics of SWNT quantum dots.

\subsection{Model Hamiltonian}

In the following we examine the physics of a generic quantum dot,
i.e. we consider a small metallic system weakly coupled to a source
and a drain via tunnelling junctions. Moreover the electrochemical
potential $\mu_{g}$ within the dot can be controlled by a capacitatively
coupled gate voltage $V_{g}$. We describe the overall system by the Hamiltonian\begin{equation}
H=H_{\odot}+H_{s}+H_{d}+H_{T}+H_{g},\label{eq:H_tot}\end{equation}
where $H_{\odot}$ can describe an interacting SWNT (cf. equation
(\ref{eq:Hac-diagonal}) below) or another conductor with known many-body
eigenstates. $H_{s/d}$ describe the isolated metallic source and
drain contacts as a Fermi gas of noninteracting quasi-particles. Upon
absorbing terms proportional to the external source/ drain voltage
$V_{s/d}$, they read ($l=s,d$) \begin{equation}
H_{l}=\sum_{\sigma\vec{q}}\varepsilon_{\vec{q},l}c_{\vec{q}\sigma l}^{\dagger}c_{\vec{q}\sigma l},\end{equation}
 where $c_{\vec{q}\sigma l}^{\dagger}$ creates a quasi-particle with
spin $\sigma$ and energy $\varepsilon_{\vec{q},l}=\varepsilon_{\vec{q}}-eV_{s/d}$
in lead $s/d$. The transfer of electrons between the leads and the
central system is taken into account by the tunnelling Hamiltonian\begin{equation}
H_{T}=\sum_{l=s,d}\sum_{\sigma}\int\textrm{d}^{3}r\left(T_{l}(\vec{r})\Psi_{\sigma}^{\dagger}(\vec{r})\Phi_{\sigma l}(\vec{r})+\textrm{h.c.}\right),\label{eq:H_T}\end{equation}
 where $\Psi_{\sigma}^{\dagger}(\vec{r})$ and $\Phi_{\sigma l}^{\dagger}(\vec{r})=\sum_{\vec{q}}\phi_{\vec{q}}^{*}(\vec{r})c_{\vec{q}\sigma l}^{\dagger}$
are electron creation operators in the dot and in lead $l$, respectively,
and $T_{l}(\vec{r})$ describes the transparency of the tunnelling
contact at lead $l$. Finally, $H_{g}=-\mu_{g}\mathcal{N}_{c}=-e c V_g$ accounts
for the gate voltage capacitively coupled to the dot, with $\mathcal{N}_{c}$
counting the total electron number in the dot and $c$ being a conversion factor, that relates the electrochemical potential to the gate voltage.

\subsection{Dynamics of the reduced density matrix \label{sub:Transport}\label{sub:Master-Equation}}

Our starting point is the Liouville equation for the time evolution
of the density matrix $\rho(t)$ of the system consisting of the leads
and the dot. The tunnelling Hamiltonian $H_{T}$ from equation (\ref{eq:H_T})
is treated as perturbation. We calculate the time dependence of $\rho(t)$
in the interaction picture, i.e. we define\begin{equation}
\rho^{I}(t)=U_{I}(t,t_{0})\rho(t_{0})U_{I}^{\dagger}(t,t_{0}),\label{eq:interactionpict-rho}\end{equation}
with the time evolution operator $U_{I}(t,t_{0})$, given by\begin{equation}
U_{I}(t,t_{0})=e^{\frac{i}{\hbar}\left(H_{\odot}+H_{s}+H_{d}\right)(t-t_{0})}e^{-\frac{i}{\hbar}\left(H_{\odot}+H_{s}+H_{d}+H_{T}\right)(t-t_{0})},\label{eq:UI}\end{equation}
with $t_{0}$ being some reference time. Using (\ref{eq:interactionpict-rho})
and (\ref{eq:UI}) the equation of motion becomes\begin{equation}
i\hbar\frac{\partial\rho^{I}(t)}{\partial t}=\left[H_{T}^{I}(t),\rho^{I}(t)\right],\label{eq:liouville-diff}\end{equation}
with $H_{T}^{I}(t)=e^{\frac{i}{\hbar}\left(H_{\odot}+H_{s}+H_{d}\right)(t-t_{0})}H_{T}e^{-\frac{i}{\hbar}\left(H_{\odot}+H_{s}+H_{d}\right)(t-t_{0})}.$
Equivalently we can write\begin{equation}
\rho^{I}(t)=\rho^{I}(t_{0})-\frac{i}{\hbar}\int_{t_{0}}^{t}\left[H_{T}^{I}(t_{1}),\rho(t_{1})\right]dt_{1}.\label{eq:liouville-int}\end{equation}
Reinserting (\ref{eq:liouville-int}) back into (\ref{eq:liouville-diff})
yields \begin{multline}
\dot{\rho^{I}}(t)=-\frac{i}{\hbar}\left[H_{T}^{I},\rho(t_{0})\right]\\
+\left(\frac{i}{\hbar}\right)^{2}\int_{t_{0}}^{t}dt_{1}\left[H_{T}^{I}(t),\left[H_{T}^{I}(t_{1}),\rho^{I}(t_{1})\right]\right].\label{eq:liouville_interaction_repr}\end{multline}
Since we are interested in the transport through the central system,
it is sufficient to consider the RDM $\rho_{\odot}^{I}$ of the dot,
which can be obtained from $\rho^{I}$ by tracing out the lead degrees
of freedom, i.e. \begin{equation}
\rho_{\odot}^{I}=Tr_{\mathsf{leads}}\rho^{I}.\label{eq:lead_trace}\end{equation}
In general the leads can be considered as large systems compared to
the dot. Besides we only consider the case of weak tunnelling, such
that the influence of the central system on the leads is only marginal.
Thus from now on we treat the leads as reservoirs which stay in thermal
equilibrium and make the following ansatz to factorise the density
matrix  $\rho^{I}(t)$ of the total system as \begin{equation}
\rho^{I}(t)=\rho_{\odot}^{I}(t)\rho_{s}\rho_{d}=:\rho_{\odot}^{I}(t)\rho_{\mathsf{leads}},\label{densitymatrix_factorized}\end{equation}
where $\rho_{s}$ and $\rho_{d}$ are time independent and given by
the usual thermal equilibrium expression\begin{equation}
\rho_{s/d}=\frac{e^{-\beta(H_{s/d}-\mu_{s/d}\mathcal{N}_{s/d})}}{Z_{s/d}},\label{eq:lead-thermalized}\end{equation}
with $\beta$ the inverse temperature. As it can be formally shown \cite{Blum1996},
the factorization (\ref{densitymatrix_factorized}) corresponds, like
Fermi's Golden Rule, to a second order treatment in the perturbation
$H_{T}.$ Furthermore, we can significantly simplify equation (\ref{eq:liouville_interaction_repr})
by making the so called Markov approximation (MA). The idea is that
the dependence of $\dot{\rho}^{I}(t)$ on $\rho^{I}(t')$ is only
local in time. In more detail, $\rho^{I}(t')$ is replaced by $\rho^{I}(t)$
in (\ref{eq:liouville_interaction_repr}). The MA is closely connected
to the correlation time $\tau$ of the leads. In our case $\tau$
is the time after which the correlation functions 
\begin{multline*}
\left\langle \Phi_{\sigma l}^{\dagger}(\vec{r},\tau)\Phi_{\sigma l}(\vec{r}',0)\right\rangle _{\mathsf{th}}:=\\ Tr_{\mathsf{leads}}\left(\Phi_{\sigma l}^{\dagger}(\vec{r},\tau)\Phi_{\sigma l}(\vec{r}',0)\rho_{\mathsf{leads}}\right)
\end{multline*}
are vanishing and therefore erasing the memory of the system. If $\rho^{I}(t')$
is not considerably changing during $\tau$ the MA is valid. It must
be noted that the MA leads to an averaging of the time evolution of
$\rho^{I}(t)$ on timescales of the order of $\tau$, such that details
of the dynamics on short time scales are not accessible. Since we
are interested in the dc current through the system, this imposes
no restriction on our purpose. Finally we get, by plugging equations
(\ref{eq:lead_trace}), (\ref{densitymatrix_factorized}) and (\ref{eq:lead-thermalized})
into (\ref{eq:liouville_interaction_repr}), the following expression
for the equation of motion for the RDM,\begin{multline}
\dot{\rho_{\odot}^{I}}(t)=-\frac{i}{\hbar}Tr_{\mathsf{leads}}\left[H_{T}^{I},\rho_{\odot}^{I}(t_{0})\rho_{\mathsf{leads}}\right]\\
+\left(\frac{i}{\hbar}\right)^{2}Tr_{\mathsf{leads}}\int_{t_{0}}^{t}dt_{1}\left[H_{T}^{I}(t),\left[H_{T}^{I}(t_{1}),\rho_{\odot}^{I}(t)\rho_{\mathsf{leads}}\right]\right].\label{eq:liouville-start}\end{multline}
The first term vanishes because 
%\begin{equation*}
$\left<\Phi_{\sigma l}(\vec{r})\right>_{\mathrm{th}}=0.$
%\end{equation*}
Since we are only interested in the longterm behavior of the system
we send $t_{0}\rightarrow-\infty.$ Writing out the double commutator
in (\ref{eq:liouville-start}) according to \[
\left[A,\left[B,C\right]\right]=ABC+CBA-ACB-BCA,\]
 and introducing the variable $t'=t-t_{1}$ one obtains\begin{multline}
\dot{\rho_{\odot}^{I}}(t)=\\
-\frac{1}{\hbar^{2}}Tr_{\mathsf{leads}}\int_{0}^{\infty}dt'\left[\left(H_{T}^{I}(t)H_{T}^{I}(t-t')\rho_{\odot}^{I}(t)\rho_{\mathsf{leads}}+h.c.\right)\right.\\
-\left.\left(H_{T}^{I}(t)\rho_{\odot}^{I}(t)\rho_{\mathsf{leads}}H_{T}^{I}(t-t')+h.c.\right)\right].\label{eq:liouville1}\end{multline}
Now we insert the explicit form of $H_{T}$ from equation (\ref{eq:H_T})
into (\ref{eq:liouville1}) and perform the trace over the lead degrees
of freedom. Thereby we exploit the relations \[
\left<\Phi_{\sigma l}\Phi_{\sigma l'}\right>_{\mathrm{th}}=\left<\Phi_{\sigma l}^{\dagger}\Phi_{\sigma l'}^{\dagger}\right>_{\mathrm{th}}=0,\]
 as well as \[
\left<\Phi_{\sigma l}\Phi_{\sigma'l'}^{\dagger}\right>_{\mathrm{th}}=0\quad\mathrm{for}\,\sigma l\neq\sigma'l',\]
 such that \begin{multline}
\dot{\rho}_{\odot}^{I}(t)=-\frac{1}{\hbar^{2}}\sum_{l=s,d}\sum_{\sigma}\int\textrm{d}^{3}x\int\textrm{d}^{3}y\int_{0}^{\infty}dt'\\
\left\{\left[\mathcal{E}_{\sigma l}(\vec{x},\vec{y},t')\Psi_{\sigma}^{\dagger}(\vec{x},t)\Psi_{\sigma}(\vec{y},t-t')\rho_{\odot}^{I}(t)\right.\right.\\
+\left.\mathcal{F}_{\sigma l}(\vec{x},\vec{y},t')\Psi_{\sigma}(\vec{x},t)\Psi_{\sigma}^{\dagger}(\vec{y},t-t')\rho_{\odot}^{I}(t)\right]+h.c.\\
-\left[\mathcal{F}_{\sigma l}^{*}(\vec{x},\vec{y},t')\Psi_{\sigma}^{\dagger}(\vec{x},t)\rho_{\odot}^{I}(t)\Psi_{\sigma}(\vec{y},t-t')\right.\\
+\left.\left.\mathcal{E}_{\sigma l}^{*}(\vec{x},\vec{y},t')\Psi_{\sigma}(\vec{x},t)\rho_{\odot}^{I}(t)\Psi_{\sigma}^{\dagger}(\vec{y},t-t')\right]-h.c.\right\}.\label{eq:GMEtracedout}\end{multline}
In (\ref{eq:GMEtracedout}) we have introduced
\begin{multline}
\mathcal{E}_{\sigma l}(\vec{x},\vec{y},t'):=T_{l}(\vec{x})T_{l}^{*}(\vec{y})\left\langle \Phi_{\sigma l}(\vec{x})\Phi_{\sigma l}^{\dagger}(\vec{y},-t')\right\rangle _{\mathrm{th}}=\\
T_{l}(\vec{x})T_{l}^{*}(\vec{y})\int d\varepsilon\rho_{l}^{\ominus}(\varepsilon)\sum_{\vec{q}|_{\varepsilon}}\phi_{\vec{q}}(\vec{x})\phi_{\vec{q}}^{*}(\vec{y})e^{-\frac{i}{\hbar}(\varepsilon-eV_{l})t'},\label{CalE_sigma_l}\end{multline}
with $\rho_{l}^{\ominus}(\varepsilon)=\rho_{l}(\varepsilon)(1-f(\varepsilon))$,
where $\rho_{l}(\varepsilon)$ is the density of energy levels in lead
$l$ and $f(\varepsilon)$ is the Fermi distribution. Similarly\begin{multline}
\mathcal{F}_{\sigma l}(\vec{x},\vec{y},t'):=T_{l}^{*}(\vec{x})T_{l}(\vec{y})\left\langle \Phi_{\sigma l}^{\dagger}(\vec{x})\Phi_{\sigma l}(\vec{y},-t')\right\rangle _{\mathrm{th}}=\\
T_{l}^{*}(\vec{x})T_{l}(\vec{y})\int d\varepsilon\rho_{l}^{\oplus}(\varepsilon)\sum_{\vec{q}|_{\varepsilon}}\phi_{\vec{q}}^{*}(\vec{x})\phi_{\vec{q}}(\vec{y})e^{\frac{i}{\hbar}(\varepsilon-eV_{l})t'}.\label{eq:CalF_sigma_l}\end{multline}
Here $\rho_{l}^{\oplus}(\varepsilon)=\rho_{l}(\varepsilon)f(\varepsilon).$
In order to proceed, it is convenient to represent the RDM in the
eigenstate basis of the dot Hamiltonian $H_{\odot}.$ Assuming that
we can diagonalize the many-body Hamiltonian $H_{\odot}$ (see Section \ref{sub:Low-energy-description}), this allows us to extract the $t$ and $t'$ dependence
of the electron operators in (\ref{eq:GMEtracedout}). To proceed
further, we carry out the following two major approximations:

I.) We assume that matrix elements between states representing different
charge states vanish (the number of electrons in the dot influences
the electrostatics of the whole circuit, hence is {}``measured''
permanently). 

II.) The so called secular approximation is applied, i.e. we only
retain those terms of (\ref{eq:GMEtracedout}), which have no oscillatory
behavior in $t.$ The latter causes that coherences between \textit{non}degenerate
states are not present in the stationary solution of (\ref{eq:GMEtracedout}).
For the dynamics this means that we can not resolve the evolution of
$\rho_{\odot}^{I}(t)$ on time scales of $\hbar/(E_{m}-E_{n}),$ where
$E_{m}$ and $E_{n}$ are two distinct energy levels of $H_{\odot}.$

In the end we can divide $\rho_{\odot}^{I}(t)$ into block matrices
$\rho_{\odot}^{IE_{N}}(t),$ where $E_{N}$ denotes an energy level
of the isolated dot containing $N$ electrons. To simplify the notation
we give the resulting equations of motion in Bloch-Redfield form,\begin{multline}
\dot{\rho}_{nm}^{I,E_{N}}(t)=-\sum_{kk'}R_{nm\, kk'}^{E_{N}}\rho_{kk'}^{I,E_{N}}(t)\\
+\sum_{M=N\pm1}\sum_{E'}\sum_{kk'}R_{nm\, kk'}^{E_{N}\, E'_{M}}\rho_{kk'}^{I,E'_{M}}(t),\label{eq:generalized_Mequ}\end{multline}
where the indices $n,m,k,k'$ refer to the eigenstates of $H_{\odot}.$
The Redfield tensors are given by $(l=s,d)$\begin{equation}
R_{nm\, kk'}^{E_{N}}=\sum_{l}\sum_{M,E',j}\left(\delta_{mk'}\Gamma_{l,njjk}^{(+)E_{N}\, E'_{M}}+\delta_{nk}\Gamma_{l,k'jjm}^{(-)E_{N}\, E'_{M}}\right),\label{eq:Blochredf_1}\end{equation}
 and \begin{equation}
R_{nm\, kk'}^{E_{N}\, E'_{M}}=\sum_{l,\alpha=\pm}\Gamma_{l,k'mnk}^{(\alpha)E'_{M}\, E_{N}}.\label{eq:Blochredf_2}\end{equation}
The quantities $\Gamma_{l,k'mnk}^{(\alpha)E_{N}\, E'_{M}}$ determine
the transitions between states with $N$ particles and energy $E_{N}$
to states with $M$ particles and energy $E'_{M}.$ In detail we obtain
for transitions $N\rightarrow N+1$\begin{multline}
\Gamma_{l\, k'mnk}^{(\alpha)E_{N}\, E'_{N+1}}=\frac{1}{\hbar^{2}}\sum_{\sigma}\int d^{3}x\int d^{3}y\times\\
\left(\Psi_{\sigma}(\vec{x})\right)_{k'm}^{E_{N}\, E'_{N+1}}\left(\Psi_{\sigma}^{\dagger}(\vec{y})\right)_{nk}^{E'_{N+1}\, E_{N}}\times\\
\int_{0}^{\infty}dt'\mathcal{F}_{\sigma l}(\vec{x},\vec{y},t')e^{-\alpha\frac{i}{\hbar}\left(E'_{N+1}-E_{N}\right)t'}.\label{eq:gamma_lEnEnp1}\end{multline}
For transitions $N\rightarrow N-1$ is \begin{multline}
\Gamma_{l\, k'mnk}^{(\alpha)E_{N}\, E_{N-1}}=\frac{1}{\hbar^{2}}\sum_{\sigma}\int d^{3}x\int d^{3}y\times\\
\left(\Psi_{\sigma}^{\dagger}(\vec{x})\right)_{k'm}^{E_{N}\, E'_{N-1}}\left(\Psi_{\sigma}(\vec{y})\right)_{nk}^{E'_{N-1}\, E_{N}}\times\\
\int_{0}^{\infty}dt'\mathcal{E}_{\sigma l}(\vec{x},\vec{y},t')e^{-\alpha\frac{i}{\hbar}\left(E'_{N-1}-E_{N}\right)t'},\label{eq:gamma_lEnEnm1}\end{multline}
where we have defined the matrix elements \begin{equation} \left(\Psi_{\sigma}^{\dagger}(\vec{x})\right)_{km}^{E_{N}\, E'_{N+1}}:=\left\langle k\left|\Psi_{\sigma}^{\dagger}(\vec{x})\right|m\right\rangle,
\end{equation}
with the states $\left|k\right\rangle $ and $\left|m\right\rangle $
having energy $E_{N},\, E'_{N+1}$ and particle number $N,\, N+1$
respectively. 

Equation (\ref{eq:generalized_Mequ}) governs the dynamics of the
SWNT electrons. In the following we deduce therefrom the current through
the system.

\subsection{Current\label{sub:Current}}

The current is essentially the net tunnelling rate in a certain direction
at one of the leads. Thus the current at lead $l$ will be of the
form \begin{equation}
I_{l}=le\sum_{N}\left(\Sigma_{l}^{N\rightarrow N+1}-\Sigma_{l}^{N\rightarrow N-1}\right),\label{eq:CurrentGeneral}\end{equation}
where we use the convention $l=s/d=\pm1.$ On the long run the currents
at the two leads have to be equal, otherwise charge would accumulate
on the dot which is prevented by the charging energy. The rates $\Sigma_{l}^{N\rightarrow N\pm1}$
can be obtained from the time evolution of the occupation probabilities
$P_{N}=Tr\left(\rho_{\odot}^{I,N}\right)$, where $\rho_{\odot}^{I,N}$
is the RDM for states containing $N$ electrons. In more detail, the
occupation probability of the charge state $N$ is reduced by tunnelling events
changing the number of electrons from $N$ to $N\pm1$ and is increased
by processes transferring the charge states $N\pm1$ to $N,$ hence
\begin{multline*}
\dot{P}_{N}=Tr\left(\dot{\rho}_{\odot}^{I,N}\right)=\\
\sum_{l}\left(-\Sigma_{l}^{N\rightarrow N+1}-\Sigma_{l}^{N\rightarrow N-1}+\Sigma_{l}^{N+1\rightarrow N}+\Sigma_{l}^{N-1\rightarrow N}\right).\end{multline*}
Using $Tr\left(\dot{\rho}_{\odot}^{I,N}\right)=\sum_{E}Tr\left(\dot{\rho}_{\odot}^{I,E_{N}}\right)$
together with (\ref{eq:generalized_Mequ}) we can easily identify
the rates appearing in (\ref{eq:CurrentGeneral}):\begin{multline}
\Sigma_{l}^{N\rightarrow N\pm1}=\label{eq:NNp1rates}\\
\sum_{E,E'}\sum_{nkj}\left(\Gamma_{l,njjk}^{(+)E_{N}\, E'_{N\pm1}}\rho_{kn}^{I,E_{N}}+\rho_{nk}^{I,E_{N}}\Gamma_{l,kjjn}^{(-)E_{N}\, E'_{N\pm1}}\right).\end{multline}
Since the relation $\Gamma_{l,njjk}^{(+)E_{N}\, E'_{N\pm1}}=\left(\Gamma_{l,kjjn}^{(-)E_{N}\, E'_{N\pm1}}\right)^{*}$
holds, we arrive, by inserting (\ref{eq:NNp1rates}) into (\ref{eq:CurrentGeneral}),
at the following expression,\begin{multline}
I_{l}=\\
l2e\mathrm{Re}\sum_{N,E,E'}\left(\Gamma_{l,njjk}^{(+)E_{N}\, E'_{N+1}}-\Gamma_{l,njjk}^{(+)E_{N}\, E'_{N-1}}\right)\rho_{kn\ }^{I,E_{N}}.\label{eq:Current2}\end{multline}
For the actual calculations we are going to replace $ \rho^{I,E_{N}}$ in (\ref{eq:Current2}) by the stationary solution $\rho_{st}^{I,E_{N}}$ of (\ref{eq:GMEtracedout}), because we are only interested in the longterm behavior of the system.

Until now our treatment has been quite general, having made no assumptions
about the nature of the dot so far. But the actual transport properties
depend on the microscopic structure of the system. As seen from equations
(\ref{eq:gamma_lEnEnp1}) and (\ref{eq:gamma_lEnEnm1}), we shall
have to determine the spectrum of $H_{\odot}$ and the matrix elements
$\left(\Psi_{\sigma}(\vec{x})\right)_{kk'}^{E_{N}\, E'_{N\pm1}}$
in order to solve master equation (\ref{eq:generalized_Mequ}). This is the subject of the next section. Moreover
the leads and the geometry of the tunnelling contacts will influence
the system via the quantities $\mathcal{F}_{\sigma l}$ and $\mathcal{E}_{\sigma l}$ as discussed in 
Section \ref{sub:Influenceofthe} and Appendix \ref{sec:Description-of-the}.

\section{Low energy description of metallic finite size SWNTs\label{sub:Low-energy-description}}

In the following we derive the Hamiltonian $H_{\odot}$ describing
the interacting electrons in a metallic finite size SWNT at low energies.
Typically (depending on the diameter of the nanotube) {}``low energies''
mean a range of 1 eV around the Fermi energy, such that the dispersion
relation for the noninteracting electrons is linear \cite{Egger1997}
around the Fermi points. Since we are interested in the transport
across a finite size system, open boundary conditions (OBCs) for the
wave functions must be imposed and we can deduce the wave functions
and the energy spectrum of the noninteracting system. Switching on
interactions we still can diagonalize the Hamiltonian by using the
so called constructive bosonization procedure \cite{Haldane1981,Delft1998},
which we also employ in order to rewrite the electron operators in
a suitable form for the determination of the matrix elements $\left(\Psi_{\sigma}(\vec{x})\right)_{kk'}^{E_{N}\, E'_{N+1}}$.

\subsection{Noninteracting SWNTs \label{sub:Noninteracting-SWNT-Hamiltonian}}

\begin{figure}
\includegraphics[%
  bb=0bp 22bp 319bp 142bp,
  clip,
  width=1.0\columnwidth]{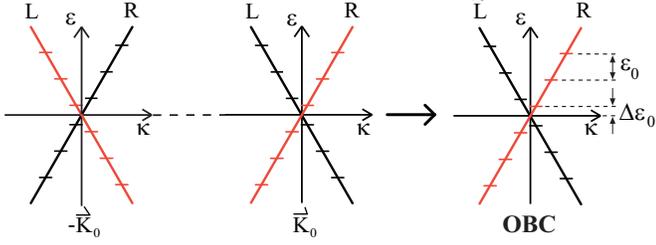}

\caption{\label{cap:Energy-spectrum-of} On the right, energy spectrum of
a SWNT with open boundary conditions (OBCs) described in terms of
left ($\tilde{L}$) and right ($\tilde{R}$) branches. It is constructed
from suitable combinations of travelling waves (cf. equation (\ref{eq:OBCfromPBC}))
whose spectrum is shown on the left side. }
\end{figure}

The bandstructure of SWNTs is conveniently derived from the one of
the unbound $2p_{z}$ electrons in graphene sheets \cite{Saito1998}.
Since each unit cell of the graphene lattice contains two atoms $p=1,2$,
the band structure consists of a valence and a conduction band which
touch at the corner points of the first Brillouin zone. Two of those
Fermi points, $F=\pm\vec{K}_{0}$, are independent, i.e. don't differ
by a reciprocal lattice vector. The energy dispersion around the Fermi
points is very well linearly approximated with a Fermi velocity $v_{F}=8.1\cdot10^{5}\, m/s$.
Since SWNTs are essentially graphene sheets rolled up to a cylinder
of a certain diameter, we obtain the SWNT band structure by imposing
periodic boundary conditions around the circumference $L_{\perp}$
of the tube, leading to the quantization of the allowed transverse
wave vectors \begin{equation}
k_{\perp}=\frac{2\pi}{L_{\perp}}m,\quad m=0,\pm1,\pm2\dots\quad,\label{eq:QuantCondPerp}\end{equation}
 and thus to the formation of subbands labelled by $m.$ For metallic
SWNTs the Fermi points $F$ will satisfy condition (\ref{eq:QuantCondPerp}),
hence the corresponding valence and conduction subbands will have
no gap. In the following the focus is on armchair SWNTs. Then only
the gapless sub-bands with linear energy dispersion nearby the Fermi
points are relevant \cite{Egger1997,Kane1997}. At each Fermi point
there are two different branches $r=R/L$ associated to right and
left moving electrons. The corresponding Bloch waves are of the form
\begin{equation}
\varphi_{R/L,F,\kappa}(\vec{r})=e^{i\kappa x}\varphi_{R/L,F}(\vec{r}),\label{eq:travellingwaves}\end{equation}
 where $\kappa$ measures the distance from the Fermi points $\pm\vec{K}_{0}$
(Fig. \ref{cap:Energy-spectrum-of} left).

In more detail, the Bloch waves $\varphi_{R/L,F}(\vec{r})$ at the
Fermi points are given by a superposition of wave functions living
either on sublattice $p=1$ or $p=2,$\begin{equation}
\varphi_{R/L,F}(\vec{r})=\frac{1}{\sqrt{N_{L}}}\sum_{\vec{R},p}e^{iF\cdot\vec{R}}f_{prF}\chi(\vec{r}-\vec{R}-\vec{\tau}_{p}),\label{eq:sublatticewaves}\end{equation}
where $N_{L}$ is the number of lattice sites which are identified
by the lattice vector $\vec{R}.$ The position of the atoms $p=1,\ 2$
in the unit cell is given by $\tau_{p}$ and the functions $\chi$
are the $p_{z}$ orbitals. The value of the coefficients $f_{prF}$
depends on the considered SWNT type. For simplicity we concentrate
on armchair SWNTs for which we have $f_{1rF}=1/\sqrt{2},\, f_{2rF}=-1/\sqrt{2}\mathrm{sgn}(rF),$
where we use the convention that $R/L=\pm1.$ In this article,
we are interested in finite size effects and therefore we impose OBCs
instead of periodic boundary conditions on the single electron wave functions. Generalizing \cite{Fabrizio1995}
to the case of SWNTs we introduce standing waves which fulfil the
OBCs (Fig. \ref{cap:Energy-spectrum-of} right): \begin{equation}
\varphi_{\tilde{R}/\tilde{L},\kappa}^{OBC}(\vec{r})=\frac{1}{\sqrt{2}}\left[\varphi_{R/L,K_{0},\kappa}(\vec{r})-\varphi_{L/R,-K_{0},-\kappa}(\vec{r})\right]\;,\label{eq:OBCfromPBC}\end{equation}
 with quantization condition \[
\kappa=\pi(m_{\kappa}+\Delta)/L,\quad m_{\kappa}=0,\pm1,\pm2,\dots,\]
where $L$ is the SWNT length. The offset parameter $\Delta$ occurs
if there is no integer $n$ with $\vec{K}_{0}=(\pi n/L)\hat{e}_{\parallel}$ (where $\hat{e}_{\parallel}$ is the unit vector along the tube axis) and is responsible for a possible energy mismatch between the states of the 
$\tilde{R}$ and $\tilde{L}$ branches, defined by relation (\ref{eq:OBCfromPBC}).
From the linear energy dispersion relation around the Fermi points,
the energies of the standing waves follow as \begin{equation}
\varepsilon_{\tilde{r}\kappa}=\mathrm{sgn}(\tilde{r})\hbar v_{F}\kappa.\label{eq:linearenergydisp}\end{equation}

Including the spin degree of freedom, the electron operator reads
\begin{equation}
\Psi(\vec{r})=\sum_{\tilde{r}=\tilde{R},\tilde{L}}\sum_{\kappa,\sigma}\varphi_{\tilde{r}\kappa}^{OBC}(\vec{r})c_{\tilde{r}\sigma\kappa}=:\sum_{\sigma}\Psi_{\sigma}(\vec{r}),\label{elektronopdef}\end{equation}
where the operator $c_{\tilde{r}\sigma\kappa}$ annihilates the state
$\left|\varphi_{\tilde{r}\kappa}^{OBC}\right\rangle \left|\sigma\right\rangle $.
Since the Bloch waves can be divided into a slowly and a fast oscillating
part (cf. (\ref{eq:travellingwaves})) we can as well split off a
slowly varying part from the operators $\Psi_{\sigma}(\vec{r}),$
i.e. we introduce the 1D operators \begin{equation}
\psi_{\tilde{r}\sigma F}(x)=\frac{1}{\sqrt{2L}}\sum_{\kappa}e^{i\mathrm{sgn}(F)\kappa x}c_{\tilde{r}\sigma\kappa},\end{equation}
in terms of which the 3D electron operators read

\begin{multline}
\Psi_{\sigma}(\vec{r})=\sqrt{L}\sum_{F}\mathrm{sgn}(F)\times\\
\left[\varphi_{\mathrm{sgn}(F)R,F}(\vec{r})\psi_{\tilde{R}\sigma F}(x)+\varphi_{\mathrm{sgn}(F)L,F}(\vec{r})\psi_{\tilde{L}\sigma F}(x)\right].\label{eq:Psisigmaexp}\end{multline}
Later on, the bosonizability of $\psi_{\tilde{r}\sigma F}(x)$ will
be of quite some significance for the calculation of the transport
properties. 

From (\ref{eq:linearenergydisp}) the appropriate Hamiltonian for
the noninteracting system is easily derived. It reads\begin{equation}
H_{\odot,0}=\hbar v_{F}\sum_{\tilde{r}\sigma}\mathrm{sgn}(r)\sum_{\kappa}\kappa c_{\tilde{r}\sigma\kappa}^{\dagger}c_{\tilde{r}\sigma\kappa}.\label{eq:NoninteractingH}\end{equation}

\subsection{Interacting SWNTs\label{sub:Interacting-SWNT-Hamiltonian}}

\begin{figure}
\includegraphics[%
  bb=25bp 25bp 320bp 120bp,
  clip,
  width=1.0\columnwidth]{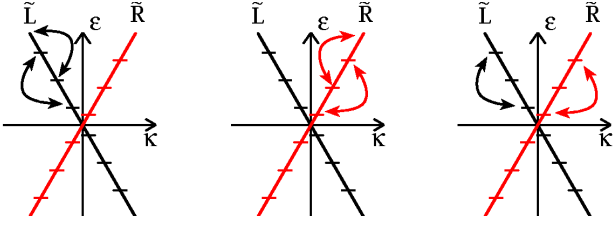}

\caption{\label{cap:forwardscattering}Scheme of forward scattering terms.
All those processes conserve the number of electrons in each branch.}
\end{figure}

For the inclusion of the electron - electron interactions we have
to add the following term to the SWNT Hamiltonian, \begin{multline}
V_{\odot}=\\
\frac{1}{2}\sum_{\sigma\sigma'}\int d^{3}r\int d^{3}r'\Psi_{\sigma}^{\dagger}(\vec{r})\Psi_{\sigma'}^{\dagger}(\vec{r}')V(\vec{r}-\vec{r}')\Psi_{\sigma'}(\vec{r}')\Psi_{\sigma}(\vec{r}),\label{eq:interactionpart}\end{multline}
where $V(\vec{r}-\vec{r}')$ is the possibly screened Coulomb potential.
Upon inserting (\ref{eq:Psisigmaexp}) into (\ref{eq:interactionpart}),
integration over the coordinates perpendicular to the tube axis yields
the interacting Hamiltonian expressed in terms of 1D operators and
an effective 1D interaction $V_{\mathrm{eff}}(x,x'),$ see (\ref{eq:Veff})
below. As in \cite{Egger1997} we expand the Bloch waves into their
sublattice contributions (\ref{eq:sublatticewaves}) and ignore the
difference between the intra and inter lattice correlations, which
are only of relevance at the length scale of the next neighbor spacing
of the carbon atoms. The final form of $V_{\odot}$ we obtain by ignoring
all fast oscillating terms in the interaction, or in other words by
keeping only the so called forward scattering processes associated
to the effective potential $V_{\mathrm{eff}}$ as depicted in Fig. \ref{cap:forwardscattering}.
This leads to an expression of $V_{\odot}$ in terms of the 1D densities
$\rho_{\tilde{r}\sigma F}(x)=\psi_{\tilde{r}\sigma F}^{\dagger}(x)\psi_{\tilde{r}\sigma F}(x),$\begin{multline}
V_{\odot}=\\
\frac{1}{2}\sum_{\tilde{r}\tilde{r}'}\sum_{FF'}\sum_{\sigma\sigma'}\int\int_{0}^{L}dxdx'\rho_{\tilde{r}\sigma F}(x)V_{\mathrm{eff}}(x,x')\rho_{\tilde{r}'\sigma'F'}(x').\label{eq:Veldens}\end{multline}
The approximations that have been made for getting from (\ref{eq:interactionpart}) to (\ref{eq:Veldens}) mainly mean that we neglect, as already mentioned in the introduction, any kind of exchange effects and thus are only valid for large enough SWNTs.
Using (\ref{eq:sublatticewaves}) and (\ref{elektronopdef})
the effective potential $V_{\mathrm{eff}}$ is obtained from an integral over
the coordinates perpendicular to the tube axis of the 3D Coulomb interaction
weighted by the $p_{z}$ orbitals, i.e.\begin{multline}
V_{\mathrm{eff}}(x,x')=\frac{L^{2}}{N_{L}^{2}}\times\\
\sum_{\vec{R},\vec{R}'}\int d^{2}r_{\perp}\int d^{2}r'_{\perp}\left|\chi(\vec{r}-\vec{R})\right|^{2}V(\vec{r}-\vec{r}')\left|\chi(\vec{r}'-\vec{R}')\right|^{2}.\label{eq:Veff}\end{multline}

\subsection{Bosonization\label{sub:Bosonization}}

Here we show how the introduction of bosonic excitations enables us
to diagonalize the SWNT Hamiltonian $H_{\odot}=H_{\odot,0}+V_{\odot}$
by recasting it into a sum of a fermionic and a bosonic part. Besides
we give the identity that expresses the electron operators in terms
of the boson operators. The connection between fermionic and bosonic
operators can be obtained from the Fourier coefficients of the electron
density operators. To be more specific, we Fourier expand the electron
density operator, 

\begin{equation}
\rho_{\tilde{r}\sigma F}(x)=\frac{1}{2L}\sum_{q}e^{i\mathrm{sgn}(F)qx}\rho_{\tilde{r}\sigma q},\label{eq:Fourierelecdens}\end{equation}
with $q=\frac{\pi}{L}n_{q},\,n_{q}\in\mathbb{Z}$ and define the
operators\begin{equation}
b_{\sigma\mathrm{sgn}(\tilde{r})q}=\frac{1}{\sqrt{n_{q}}}\rho_{\tilde{r}\sigma \mathrm{sgn}(\tilde{r})q},\quad q>0.\label{eq:defb}\end{equation}
The operators $b_{\sigma q}$ fulfil the canonical bosonic commutation
relations $\left[b_{\sigma q},b_{\sigma'q'}^{\dagger}\right]=\delta_{\sigma\sigma'}\delta_{qq'}$
as it can be shown \cite{Delft1998} by using the explicit expression \begin{equation}
b_{\sigma\mathrm{sgn}(\tilde{r})q}=\frac{1}{\sqrt{n_{q}}}\sum_{\kappa}c_{\tilde{r}\sigma\kappa}^{\dagger}c_{\tilde{r}\sigma\kappa+\mathrm{sgn}(\tilde{r})q},\quad q>0.\label{eq:bopsexpl}\end{equation}
From the previous equation we can see that the $b,\, b^{\dagger}$
operators annihilate or create collective particle hole excitations
within the branch $(\tilde{r}\sigma)$.

\subsubsection{Diagonalization of the SWNT Hamiltonian}

As for example shown in \cite{Delft1998} for a generic 1D system,
the noninteracting part of the Hamiltonian is already diagonal in
the bosonic operators. For the SWNTs in particular, we obtain from (\ref{eq:NoninteractingH})
and (\ref{eq:bopsexpl}),
\begin{multline}
H_{\odot,0}=\\
\varepsilon_{0}\sum_{\sigma}\left[\sum_{q\neq0}\left|n_{q}\right|b_{\sigma q}^{\dagger}b_{\sigma q}+\sum_{\tilde{r}}\left(\frac{1}{2}\mathcal{N}_{\tilde{r}\sigma}^{2}+\Delta\mathrm{sgn}(\tilde{r})\mathcal{N}_{\tilde{r}\sigma}\right)\right],\label{eq:H0bosonized}\end{multline}
where $\varepsilon_{0}=\hbar v_{F}\frac{\pi}{L}$ is the level spacing of the noninteracting nanotube. The first term
in (\ref{eq:H0bosonized}) describes collective particle hole excitations,
whereas the second term represents the energy cost of the shell filling
due to Pauli's principle. Specifically, $\mathcal{N}_{\tilde{r}\sigma}=\sum_{\kappa}c_{\tilde{r}\sigma\kappa}^{\dagger}c_{\tilde{r}\sigma\kappa}$
counts the number of electrons $N_{\tilde{r}\sigma}$ in the $(\tilde{r}\sigma)$-branch.
Without loss of generality we have assumed that we only have states
with a positive number of particles in each branch. 

\begin{figure}
\includegraphics[%
  clip,
  width=1.0\columnwidth]{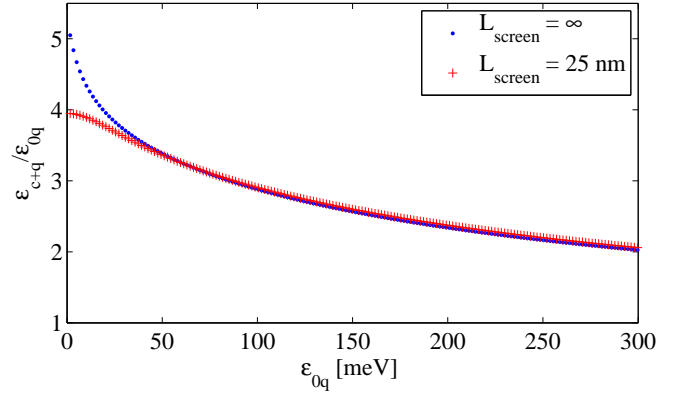}

\caption{\label{cap:ecdurche0} The ratio $\varepsilon_{c+\, q}/\varepsilon_{0\, q}$ as a function of $\varepsilon_{0\, q}$
for a (20,20) armchair SWNT of 980 nm length. Here we show the decay
of $\varepsilon_{c+\, q}/\varepsilon_{0\, q}$ for an unscreened and
a screened (screening length 25 nm) Coulomb interaction. In both cases,
a dielectric constant $\varepsilon$ of $1.4$, see Ref. \cite{Egger1997},
is assumed.}
\end{figure}

Plugging the Fourier expansion of the density operators (\ref{eq:Fourierelecdens})
into (\ref{eq:Veldens}) and taking into account the definition of
the $b$-operators yields the bosonized form of $V_{\odot},$\begin{multline}
V_{\odot}=\frac{1}{2}E_{c}\mathcal{N}_{c}^{2}+\frac{1}{2}\sum_{q>0}n_{q}W_{qq}\times\\
\sum_{\sigma\sigma'}\sum_{\tilde{r}\tilde{r}'}\left(b_{\sigma\mathrm{sgn}(\tilde{r})q}+b_{\sigma\mathrm{sgn}(\tilde{r})q}^{\dagger}\right)\left(b_{\sigma'\mathrm{sgn}(\tilde{r}')q}+b_{\sigma'\mathrm{sgn}(\tilde{r}')q}^{\dagger}\right).\label{eq:Vbosonized}\end{multline}
Here $E_{c}=W_{00}$ is the SWNT charging energy responsible for the
Coulomb blockade and the effective interaction potential is absorbed
into\begin{equation}
W_{qq}=\frac{1}{L^{2}}\int_{0}^{L}dx\int_{0}^{L}dx'V_{\mathrm{eff}}(x,x')\cos(qx)\cos(qx').\label{eq:Wqq}\end{equation}
 Since the bosonic operators appear only quadratically in (\ref{eq:H0bosonized})
and (\ref{eq:Vbosonized}), the Bogoliubov transformation \cite{Avery1976}
can be applied to diagonalize the bosonic part of the total SWNT Hamiltonian.
Here we only give the result, namely\begin{eqnarray}
H_{\odot} & = & \frac{1}{2}E_{c}\mathcal{N}_{c}^{2}+\varepsilon_{0}\sum_{\tilde{r}\sigma}\left(\frac{\mathcal{N}_{\tilde{r}\sigma}^{2}}{2}+\Delta\textrm{sgn}(\tilde{r})\mathcal{N}_{\tilde{r}\sigma}\right)\nonumber \\
 & + & \sum_{q>0}\sum_{j=c,s}\sum_{\delta=\pm}\varepsilon_{j\delta q}a_{j\delta q}^{\dagger}a_{j\delta q}.\label{eq:Hac-diagonal}\end{eqnarray}
The first line of (\ref{eq:Hac-diagonal}) describes
the energy cost to add new particles to the system. The excitations
are created by the bosonic operators $a_{j\delta q}^{\dagger}$. Four
channels are associated to total ($j\delta=c+$, $s+$) and relative
$(j\delta=c-,s-)$ (with respect to the occupation of the $\tilde{R}$
and $\tilde{L}$ branch) charge and spin excitations. To be more precise,
the new operators $a_{j\delta q}$ are related to the old operators
$b_{\sigma\mathrm{sgn}(\tilde{r})q}$ via the Bogoliubov transformation:\begin{equation}
b_{\sigma\mathrm{sgn}(\tilde{r})q}=\sum_{j\delta}\Lambda_{\tilde{r}\sigma}^{j\delta}\left(S_{j\delta q}a_{j\delta q}+C_{j\delta q}a_{j\delta q}^{\dagger}\right),\label{eq:bintermsofas}\end{equation}
where $\Lambda_{\sigma\tilde{r}}^{j\delta}$ is given by
\begin{equation}
\Lambda_{\tilde{r}\sigma}^{j\delta}=\frac{1}{2}\left(\begin{array}{cccc}
1 & 1 & 1 & 1\\
1 & 1 & -1 & -1\\
1 & -1 & 1 & -1\\
1 & -1 & -1 & 1\end{array}\right),\quad\begin{array}{cc}
j\delta= & c+,c-,s+,s\\
\tilde{r}\sigma= & \tilde{R}\uparrow,\tilde{R}\downarrow,\tilde{L}\uparrow,\tilde{L}\downarrow\end{array}.\label{eq:Lambdajdelta_sigmar}\end{equation}
For the transformation indices $S_{j\delta q}$ and $C_{j\delta q}$
we get in the case of the three modes $j\delta=c-,s+,s-$, \begin{equation}
S_{j\delta q}=1\textrm{ and }C_{j\delta q}=0.\label{eq:Sneutral}\end{equation}
 Only for $j\delta=c+$ there is an interaction dependence, \begin{equation}
S_{c+q}=\frac{1}{2}\left(\sqrt{\frac{\varepsilon_{0q}}{\varepsilon_{cq}}}+\sqrt{\frac{\varepsilon_{cq}}{\varepsilon_{0q}}}\right),\, C_{c+q}=\frac{1}{2}\left(\sqrt{\frac{\varepsilon_{0q}}{\varepsilon_{cq}}}-\sqrt{\frac{\varepsilon_{cq}}{\varepsilon_{0q}}}\right).\label{eq:Scplus}\end{equation}
 Generalized spin-charge separation occurs, since for the three interaction
independent channels $j\delta=c-,s+,s-$ the energy dispersion is
the same as for the noninteracting system, \[
\varepsilon_{j\delta q}=\hbar v_{F}q=\hbar v_{F}\frac{\pi}{L}n_{q}=:\varepsilon_{0q},\quad n_{q}=1,2,\dots\quad,\]
but the energies of the $c+$ channel are enhanced by the repulsive
interaction, \begin{equation}
\varepsilon_{c+\, q}=\varepsilon_{0q}(1+8W_{qq}/\varepsilon_{0})^{1/2}.\label{eq:dispchargedmode}\end{equation}
The ratio between $\varepsilon_{c+\, q}$ and $\varepsilon_{0\, q}$
for a (20,20)-armchair SWNT is shown in Fig. (\ref{cap:ecdurche0}).
For the actual calculation of $\varepsilon_{c+\, q}$ we have substituted
$\left|\chi(\vec{r}-\vec{R})\right|^{2}$ by $\delta(\vec{r}-\vec{R})$
in (\ref{eq:Veff}). The extension of the $p_{z}$ orbitals has been
modeled by introducing the average distance $a_{\perp}\approx0.15$
nm of the electrons from their nuclei \cite{Egger1997}, i.e. we used
for the 3D interaction potential the following expression\[
V(\vec{R}-\vec{R}')=\frac{e^{2}}{4\pi\varepsilon_{0}\varepsilon}\frac{e^{-\left|\vec{R}-\vec{R}'\right|/L_{\textrm{screen}}}}{\sqrt{\left(\vec{R}-\vec{R}'\right)^{2}+a_{\perp}^{2}}}.\]
 Here $L_{\textrm{screen}}$ is the screening length of the potential
and $\varepsilon_{0}\varepsilon$ is the dielectric constant. Due to the
finite range of the interaction potential we find a decay of $\varepsilon_{c+\, q}/\varepsilon_{0\, q}$
with increasing $q,$ which should be taken into account if higher
excitations in the $c+$ channel are involved in transport. In the
Luttinger liquid theory $\varepsilon_{c+\, q}/\varepsilon_{0\, q}$ 
is usually assumed to be equal to the constant $1/g,$ where \begin{equation}g=(1+8W_{qq}/\varepsilon_{0})^{-1/2}, \quad q=\frac{\pi}{L}.\label{eq:g}\end{equation}  The eigenstates of the SWNT Hamiltonian in (\ref{eq:Hac-diagonal})
are \begin{equation}
\left|\vec{N},\vec{m}\right\rangle:=\prod_{q>0,j\delta}\left(m_{j\delta q}!\right)^{-1/2}\left(a_{j\delta q}^{\dagger}\right)^{m_{j\delta q}}\left|\vec{N},\vec{0}\right\rangle,\label{eq:eigenstates}\end{equation}
 where $|\vec{N},\vec{0}\rangle$ has no bosonic excitations and the vector $\vec{N}$
defines the number of electrons in each of the four branches $(\tilde{r}\sigma).$

\subsubsection{Bosonization of the electron operators}

The determination of the transport properties through the SWNT quantum
dot involves the calculation of the matrix elements of the electron
operators between the eigenstates (\ref{eq:eigenstates}). Therefor
we give the so called bosonization identity for the operators $\psi_{\tilde{r}\sigma F}(x)$,\begin{equation}
\psi_{\tilde{r}\sigma F}(x)=\frac{1}{\sqrt{1-e^{-\alpha\frac{\pi}{L}}}}\eta_{\tilde{r}\sigma}K_{\tilde{r}\sigma F}(x)e^{i\phi_{\tilde{r}\sigma F}^{\dagger}(x)+i\phi_{\tilde{r}\sigma F}(x)}.\label{eq:bosident}\end{equation}
The way (\ref{eq:bosident}) is derived can be found e.g. in \cite{Delft1998}.
Here $\alpha$ is a convergence factor needed to ensure that only
excitations with physically sensible energies are considered. It will
be set to zero at the end of the calculation. The operator $\eta_{\tilde{r}\sigma}$
is the so called Klein factor. Its main effect if acting on a state
$\left|\vec{N},\vec{m}\right\rangle $ is to annihilate a particle
in the $(\tilde{r}\sigma)$-branch, more precisely\[
\eta_{\tilde{r}\sigma}\left|\vec{N},\vec{m}\right\rangle =(-1)^{\sum_{j=1}^{\tilde{r}\sigma-1}N_{j}}\left|\vec{N}-\hat{e}_{\tilde{r}\sigma},\vec{m}\right\rangle ,\]
where we use the convention $j=\tilde{R}\uparrow,\tilde{R}\downarrow,\tilde{L}\uparrow,\tilde{L}\downarrow$ $=1,2,3,4$.
The fact that the Klein factor annihilates fermions after all, expresses
itself in the factor $(-1)^{\sum_{j=1}^{\tilde{r}\sigma-1}N_{j}}.$ 
$K_{\tilde{r}\sigma F}(x)$ yields a phase depending on the number
of electrons in the $(\tilde{r}\sigma)$ branch, in our case,\[
K_{\tilde{r}\sigma F}(x)=\frac{1}{\sqrt{2L}}e^{i\frac{\pi}{L}\mathrm{sgn}(F)(\mathrm{sgn}(\tilde{r})\mathcal{N}_{\tilde{r}\sigma}+\Delta)x}.\]
At last we have the fields $\phi_{r\sigma F}(x)$ given in terms of
the bosonic operators $b_{\sigma q},$ \[
i\phi_{r\sigma F}(x)=\sum_{q>0}\frac{e^{-\alpha q/2}}{\sqrt{n_{q}}}e^{i\mathrm{sgn}(\tilde{r}F)qx}b_{\sigma\mathrm{sgn}(\tilde{r})q}.\]

\subsubsection{The matrix elements of the electron operators}

\begin{figure}
\includegraphics[%
  width=1.0\columnwidth,
  keepaspectratio]{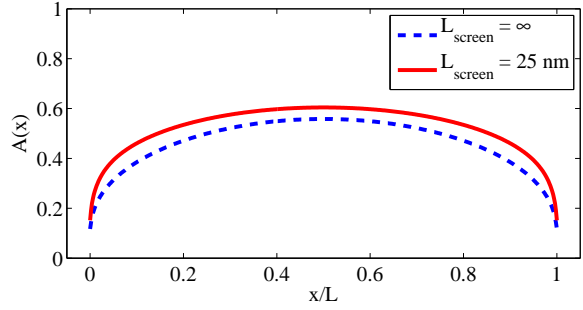}

\caption{\label{cap:W(x)-along-the}The non-oscillatory position dependence
of the matrix elements $\left\langle k\left|\psi_{\tilde{r}\sigma F}(x)\right|k'\right\rangle $
is described by $A(x),$ if no $c+$ excitations are involved. For
a repelling interaction, $A(x)$ is strongly suppressed at the SWNT
ends. Here we used the values of $\varepsilon_{cq}/\varepsilon_{0q}$ of
a (20,20) armchair SWNT ($L=980$ nm) as shown in Fig. (\ref{cap:ecdurche0}).}
\end{figure}

In the following we are going to determine the matrix elements  $\left(\Psi_{\sigma}(\vec{x})\right)_{kk'}^{E_{N}\, E'_{N\pm1}}$,
appearing in the transition rates (\ref{eq:gamma_lEnEnp1}) and (\ref{eq:gamma_lEnEnm1}).
As we know from equation (\ref{eq:Psisigmaexp}) the 3D electron operators
$\Psi_{\sigma}(\vec{x})$ can be written in terms of the 1D operators
$\psi_{r\sigma F}(x)$ yielding\begin{multline}
\left(\Psi_{\sigma}(\vec{x})\right)_{kk'}^{E_{N}\, E'_{N\pm1}}=
\sqrt{L}\sum_{F}\mathrm{sgn}(F)\times\\
\left[\varphi_{\mathrm{sgn}(F)R,F}(\vec{r})\left(\psi_{\tilde{R}\sigma F}(x)\right)_{kk'}^{E_{N}\, E'_{N\pm1}}\right.\\
+\left. \varphi_{\mathrm{sgn}(F)L,F}(\vec{r})\left(\psi_{\tilde{L}\sigma F}(x)\right)_{kk'}^{E_{N}\, E'_{N\pm1}}\right].\label{eq:Psi3Dnm}\end{multline}
Now the bosonization procedure used for the diagonalization of $H_{\odot}$
and the redefinition of $\psi_{r\sigma F}(x)$ in terms of boson operators
pays off. Denoting the SWNT eigenstates $\left|k\right\rangle $ and
$\left|k'\right\rangle $ as $\left|\vec{N},\vec{m}\right\rangle $
and $\left|\vec{N}',\vec{m}'\right\rangle $ according to equation
(\ref{eq:eigenstates}), we get, as shown in Appendix \ref{sec:The-matrix-elements},
% \begin{multline}
% \left\langle \vec{N},\vec{m}\right|\psi_{\tilde{r}\sigma F}(x)\left|\vec{N}',\vec{m}'\right\rangle =\delta_{\vec{N}+\hat{e}_{\tilde{r}\sigma},\vec{N}'}(-1)^{\sum_{j=1}^{\tilde{r}\sigma-1}N_{j}}\times\\
% K_{\tilde{r}F(\vec{N}')_{\tilde{r}\sigma}}(x)\underbrace{\frac{e^{-\frac{1}{2}\sum_{q>0}e^{-\alpha q}\sum_{j\delta}\left|\lambda_{\tilde{r}\sigmaF}^{j\deltaq}(x) \right|^{2}}}{\sqrt{1-e^{-\alpha\frac{\pi}{L}}}}}_{=:A(x)}\times\label{eq:Psinm}\\
% \prod_{q>0}\prod_{j \delta}F(\lambda_{\tilde{r} \sigma F}^{j\delta q}(x),m_{j \delta q},m'_{j \delta q}).\end{multline}
\begin{multline}
\left\langle \vec{N},\vec{m}\right|\psi_{\tilde{r}\sigma F}(x)\left|\vec{N}',\vec{m}'\right\rangle =\delta_{\vec{N}+\hat{e}_{\tilde{r}\sigma},\vec{N}'}(-1)^{\sum_{j=1}^{\tilde{r}\sigma-1}N_{j}}\times\\
K_{\tilde{r}F(\vec{N}')_{\tilde{r}\sigma}}(x)\underbrace{\frac{e^{-\frac{1}{2}\sum_{q>0}e^{-\alpha q}\sum_{j\delta}\left|\lambda_{\tilde{r}\sigma F}^{j\delta q} (x)\right|^{2}}}{\sqrt{1-e^{-\alpha\frac{\pi}{L}}}}}_{=:A(x)}\times\label{eq:Psinm}\\
\prod_{q>0}\prod_{j\delta}F(\lambda_{\tilde{r}\sigma F}^{j\delta q}(x),m_{j\delta q},m'_{j\delta q}).\end{multline}
The parameters $\lambda_{\tilde{r}\sigma F}^{j\delta q}(x)$ (cf. Appendix \ref{sec:The-parameters-}) for the
three neutral modes $j\delta=c-,\  s+,\  s-$ are given by \begin{equation}
\lambda_{\tilde{r} \sigma F}^{j \delta q}(x) =\frac{e^{i\mathrm{sgn}(F\tilde{r})qx}}{\sqrt{n_{q}}}\Lambda_{\tilde{r}\sigma}^{j\delta},\label{eq:lambdaneutral}\end{equation}
 and for the $c+$ mode we have\begin{multline}
\lambda_{\tilde{r}\sigma F}^{c+q}(x)=\\
\frac{1}{2\sqrt{n_{q}}}\left(\sqrt{\frac{\varepsilon_{cq}}{\varepsilon_{0q}}}\cos(qx)+i\sqrt{\frac{\varepsilon_{0q}}{\varepsilon_{cq}}}\mathrm{sgn}(F\tilde{r})\sin(qx)\right).\label{eq:lambdacharged}\end{multline}
The function $F(\lambda,m,m')$, describing the dependence of the matrix elements on the bosonic excitations, can be expressed in terms of the Laguerre polynomials $L_{n}^{m}$ as we show explicitly in Appendix \ref{sec:The-matrix-elements}, see equations (\ref{eq:Fexpl1}) and (\ref{eq:Fexpl2}) there.
It is interesting to note that the interaction leads to the formation
of a non-oscillatory $x$ dependence of the matrix elements $\left\langle k\left|\psi_{\tilde{r}\sigma F}(x)\right|k'\right\rangle .$
For matrix elements between states with \textit{no} bosonic $c+$
excitations this effect is described solely by the function $A(x)$.
From expressions (\ref{eq:lambdaneutral}) and (\ref{eq:lambdacharged})
for the parameters $\lambda_{\tilde{r}\sigma F}^{j\delta q}$ we find
\begin{multline}
A(x)=e^{-\sum_{q>0}\frac{e^{-\alpha q}}{8n_{q}}\left(\frac{\varepsilon_{cq}}{\varepsilon_{0q}}\cos^{2}(qx)+\frac{\varepsilon_{0q}}{\varepsilon_{cq}}\sin^{2}(qx)-1\right)}.\label{eq:Wvonxexpl}\end{multline}
Since we are considering a finite ranged interaction, the ratio $\varepsilon_{cq}/\varepsilon_{0q}$
goes to $1$ for large $q$ and so the sum in (\ref{eq:Wvonxexpl})
converges even for $\alpha=0$ (using a finite $\alpha$ in order to disregard unphysical states outside of the low energy regime has only little effect on $A(x)$). In Fig. \ref{cap:W(x)-along-the}
we show $A(x)$ for a (20,20) armchair SWNT with screened and unscreened
Coulomb potential, using the values of $\varepsilon_{cq}/\varepsilon_{0q}$ as shown in
Fig. (\ref{cap:ecdurche0}). It is evident that $A(x)$ (and hence the tunnelling amplitude at
low energies) is smallest at the tube ends and increases towards the
middle of the nanotube. For an attractive interaction the opposite
behavior would be found and for the noninteracting system $A(x)\equiv1.$
Hence $A(x)$ is closely related to the dispersion relation for the excitation energies $\varepsilon_{cq}$.
However, in this article the contact geometry is fixed and so the
actual form of $A(x)$ doesn't influence the transport properties
qualitatively, as we will show below when discussing the role of the
tunnelling contacts.

\section{\label{sub:Current-trough-a}Current through a SWNT quantum dot}

Knowing the spectrum of $H_{\odot}$ and the expressions for the electron
operator matrix elements we can return to Section \ref{sub:Transport}
and perform the actual calculation of the current through a SWNT quantum
dot.

\subsection{\label{sub:Influenceofthe}Influence of the tunnelling contacts}

Due to the integration $\int d^{3}x\int d^{3}y$, the rates from (\ref{eq:gamma_lEnEnp1})
and (\ref{eq:gamma_lEnEnm1}) depend on the geometry of the tunnelling
contact. At low enough energies (i.e. as long as no degenerate
states with very different bosonic excitations or fermionic configurations
are considered), the matrix elements of the 1D operators are slowly varying compared to the extension of the tunnelling contacts described by the transparencies $T_{l}(\vec{r}).$
Then the influence of the tunnelling contacts at the tube ends can be taken
into account by introducing the parameters (cf. Appendix \ref{sec:Description-of-the})\begin{multline}
\Phi_{l\tilde{r}\tilde{r}'}(\varepsilon)=\\
\int d^{3}r\int d^{3}r'T_{l}^{*}(\vec{r})T_{l}(\vec{r'})\sum_{\vec{q}|_{\varepsilon}}\phi_{l\vec{q}}^{*}(\vec{r})\phi_{l\vec{q}}(\vec{r'})A(x)A(x')\\
\times\sum_{FF'}\mathrm{sgn}(FF')\varphi_{\mathrm{sgn}(F)r,F}(\vec{r})\varphi_{\mathrm{sgn}(F')r',F'}^{*}(\vec{r}')\nu_{lFF'}(\Delta),\label{eq:Phirr}\end{multline}
where the phase factor $\nu_{lFF'}(\Delta)=e^{i\delta_{l,d}\pi\mathrm{sgn}(F-F')\Delta}$
accounts for the band mismatch $\Delta$. With the new parameters
$\Phi_{l\tilde{r}\tilde{r}'}(\varepsilon),$ equation (\ref{eq:gamma_lEnEnp1})
can be rewritten as \begin{multline}
\Gamma_{l\, k'mnk}^{(\alpha)E_{N}\, E'_{N+1}}=\\
\frac{1}{\hbar^{2}}\sum_{\tilde{r}\tilde{r}'\sigma}\int d\varepsilon\rho_{l}^{\oplus}(\varepsilon)\Phi_{l\tilde{r}\tilde{r}'}(\varepsilon)\left(\psi_{\tilde{r}\sigma l}\right)_{k'm}^{E_{N}\  E'_{N+1}}\times\\
\left(\psi_{\tilde{r}'\sigma l}^{\dagger}\right)_{nk}^{E'_{N+1}\, E_{N}}\int_{0}^{\infty}dt'e^{\alpha\frac{i}{\hbar}\left(\varepsilon-eV_{l}-\left(E'_{N+1}-E_{N}\right)\right)t'}\label{eq:gammaNNp1spec}\end{multline}
with $\rho_{l}^{\oplus}(\varepsilon)=\rho_{l}(\varepsilon)f(\varepsilon)$
as defined below equation (\ref{eq:CalF_sigma_l}). Additionally we
have introduced the notations $\psi_{\tilde{r}\sigma l}:=\psi_{\tilde{r}\sigma K_{0}}(x_{l})/A(x_{l})$
and $x_{l}=0,\, L$ for $l=s,\, d.$ For (\ref{eq:gamma_lEnEnm1})
the parameterization can also be performed yielding, \begin{multline}
\Gamma_{l\, k'mnk}^{(\alpha)E_{N}\, E_{N-1}}=\\
\frac{1}{\hbar^{2}}\sum_{\tilde{r}\tilde{r}'\sigma}\int d\varepsilon\rho_{l}^{\ominus}(\varepsilon)\Phi_{l\tilde{r}\tilde{r}'}^{*}(\varepsilon)\left(\psi_{\tilde{r}\sigma l}^{\dagger}\right)_{k'm}^{E_{N}\  E'_{N-1}}\times\\
\left(\psi_{\tilde{r}'\sigma l}\right)_{nk}^{E'_{N-1}\, E_{N}}\int_{0}^{\infty}dt'e^{\alpha\frac{i}{\hbar}\left(-\varepsilon+eV_{l}-\left(E'_{N-1}-E_{N}\right)\right)t'},\label{eq:gammaNNm1spec}\end{multline}
where $\rho_{l}^{\ominus}(\varepsilon)=\rho_{l}(\varepsilon)(1-f(\varepsilon))$.
The integrals over $\varepsilon$ and $t'$ in (\ref{eq:gammaNNp1spec})
and (\ref{eq:gammaNNm1spec}) can be carried out by using \begin{equation}
\int d\varepsilon g(\varepsilon)\int_{0}^{\infty}dt'e^{\alpha\frac{i}{\hbar}\left(\varepsilon-E\right)t'}=\pi\hbar g(E)+\alpha i\hbar\mathcal{P}\int\frac{g(\varepsilon)}{\varepsilon-E}d\varepsilon,\label{eq:deltaplusprincipal}\end{equation}
 where $\mathcal{P}$ denotes the Cauchy principal value. The first
term on the right hand side of (\ref{eq:deltaplusprincipal}) corresponds
to processes which conserve the energy. Additionally, we have the
principal value terms which can be attributed to so called virtual
transitions since they cancel in the expression for the current
but nevertheless can affect the transport properties indirectly via
the time evolution of the RDM. 

From now on we assume furthermore that the leads are properly described
by a 3D electron gas (think of gold leads for example). In Appendix
\ref{sec:Description-of-the} we then find for a realistic range of
the lead electron wavenumbers $q$,\begin{equation}
\Phi_{l\tilde{r}\tilde{r}'}(\varepsilon)=\delta_{\tilde{r}\tilde{r}'}\Phi_{l}(\varepsilon),\label{eq:NoPol}\end{equation}
with \begin{equation}
\Phi_{l}\sim\sum_{\vec{R},p}\left|T_{l}(\vec{R}+\vec{\tau}_{p})\right|^{2}A^{2}(\vec{R}_{x}),\label{eq:Phi3dgasexplicit}\end{equation}
simplifying equations (\ref{eq:gammaNNp1spec}) and (\ref{eq:gammaNNm1spec})
further. We should note that the case \begin{equation}
\Phi_{l\tilde{r}\tilde{r}'}(\varepsilon)\neq\delta_{\tilde{r}\tilde{r}'}\Phi_{l}(\varepsilon)\label{eq:PhiPol}\end{equation}
 corresponds to leads which are polarized with respect to the band
degree of freedom $\tilde{r}$, in complete analogy to spin polarized
leads. Hence it will be interesting to discuss in the future for which
type of contacts, (\ref{eq:PhiPol}) eventually could be fulfilled.

\subsection{\label{sub:Excitation-lines}Excitation lines}

If certain transitions are possible depends among other things on
the available energy and hence on the applied gate and bias voltages.
From (\ref{eq:gammaNNp1spec}) and (\ref{eq:gammaNNm1spec}) together
with the evaluation of the appearing $\int_{0}^{\infty}dt'\dots$
integral according to (\ref{eq:deltaplusprincipal}), we know the
resonance condition for tunnelling in/out of lead $l$. In detail,
at temperature $T=0$, transitions from a state with $N$ electrons
and eigenenergy $E_{N}$ to a state with $M=N\pm1$ electrons and
eigenenergy $E_{M}$ are possible under the condition\begin{eqnarray}
eV_{l} & \leq & E_{N}-E_{N+1}+\mu_{g},\text{ for }M=N+1,\nonumber \\
\  eV_{l} & \ge & E_{N-1}-E_{N}+\mu_{g},\text{ for }M=N-1.\label{eq:exclines}\end{eqnarray}
 We expect that for low temperatures the current can only change considerably
at the position of the corresponding lines in the $eV$ - $\mu_{g}$
plane. But we should mention that in principle the so called virtual
transitions (cf. (\ref{eq:deltaplusprincipal})) are sensitive to
the values of the bias and gate voltage outside of the excitation
lines, if coherences of the RDM influence transport. However, in our
case of unpolarized leads we do not find a significant change of the
current between two excitation lines. Furthermore not all of the resonance
conditions lead to a considerable change of the current at the corresponding
lines in the $eV_{b}$ - $\mu_{g}$ plane. Relevant transitions are
those between states with considerable overlap matrix elements $\left\langle \vec{N},\vec{m}\right|\psi_{\tilde{r}\sigma F}(x)\left|\vec{N}',\vec{m}'\right\rangle $
and for which the occupation probability of the initial state is large
enough. 

In order to add another electron to a $N$ particle ground state the
extra energy $\delta\mu_{N}=E_{N+1}^{0}-E_{N}^{0}-\left(E_{N}^{0}-E_{N-1}^{0}\right)$
has to be payed. Here $E_{M}^{0}$ denotes the ground state energies. From the SWNT Hamiltonian $H_{\odot}$, equation
(\ref{eq:Hac-diagonal}), we deduce \begin{eqnarray}
\delta\mu_{4m+1} & = & \delta\mu_{4m+3}=E_{c},\nonumber \\
\delta\mu_{4m+2} & = & E_{c}+2|\Delta|\varepsilon_{0},\nonumber \\
\delta\mu_{4m} & = & E_{c}+(1-2|\Delta|)\varepsilon_{0}.\label{eq:additionenergies}\end{eqnarray}
The energy $\delta\mu_{N}$ is a direct measure of the height and
the width of the Coulomb diamonds in the $eV_b$ - $\mu_{g}$ plane.
Thus a repeated pattern of one large Coulomb diamond followed by three
smaller ones is expected for $\Delta=0,\ 1/2$. Otherwise the pattern
will consist of a large diamond followed by a small, a medium and
again a small one. In the experiments of \cite{Sapmaz2005} sample
C showed the first pattern repeating very regularly, whereas samples
A and B revealed the second pattern. 

For sample C of \cite{Sapmaz2005} not only the Coulomb diamonds but
also a bunch of excitation lines could be resolved. In \cite{Sapmaz2005}
the positions of the Coulomb diamonds and of the lowest lying excitation
lines were determined. The mean field theory of \cite{Oreg2000} was
used for comparison. Apart from the height of the large Coulomb diamonds,
all the lines from the experimental $I-V$ characteristics of sample
C could be reproduced by an appropriate choice of five mean field
parameters $E_{c},\ \varepsilon_{0},\ \Delta,\  J$ and $dU$. The
first three parameters also appear in our theory. The mean field parameters
$J$ and $dU$ are the exchange energies with respect to the spin
and band degree of freedom. For sample C the choice of \cite{Sapmaz2005}
was $E_{c}=6.6$ meV, $\varepsilon_{0}=8.7$ meV, $2\Delta\varepsilon_{0}=J=2.9$
meV and $dU=0$ meV. With our theory exactly the same Coulomb diamonds
and excitation lines are recovered by choosing $E_{c}=9.5$ meV, $\varepsilon_{0}=2.9$
meV and $\Delta=0.$ We think that in this case our choice of parameters is much more realistic than the one made in \cite{Sapmaz2005}, with an unreasonably high $J$ of $2.9$ meV (compare this to $J=0.5$ meV in \cite{Moriyama2005}, where a considerably smaller SWNT was examined). We therefore conclude that our treatment of the interaction,
where only forward scattering events are considered and exchange
contributions are not present, is valid here. In Section
\ref{sub:High-bias-regime}, Figure \ref{cap:IDelta0} shows the numerical result for the corresponding
current.

\subsection{\label{sub:Low-Bias-Regime}Low Bias Regime}

In the following we consider the low bias regime, i.e. the bias voltages
and the temperature are low enough that only ground states with $N$
and $N+1$ particles and energies $E_{N}^{0},\, E_{N+1}^{0}$ can have
a considerable occupation probability. In this case the importance
of taking into account the off-diagonal elements of the RDM in (\ref{eq:generalized_Mequ})
depends crucially on the parameters $\Phi_{lrr'}(\varepsilon)$ from
(\ref{eq:Phirr}). From the expressions (\ref{eq:gammaNNp1spec})
and (\ref{eq:gammaNNm1spec}) for the rates it is evident that for
our assumption of unpolarized leads, i.e. under condition (\ref{eq:NoPol}),
the time evolution of the RDM elements between states with the same
band filling vector $\vec{N}$ is decoupled from elements between
states with different $\vec{N}$. Since the current only depends on
the time derivative of the diagonal elements of the RDM, the elements
mixing states with different $\vec{N}$ will have no influence on
the current and therefore can be ignored. Because all considered
ground states do have a different $\vec{N},$ in the low bias regime
we only have to take into account \textit{diagonal} matrix elements
in the master equation. %
\begin{figure}
\includegraphics[%
  width=1.0\columnwidth]{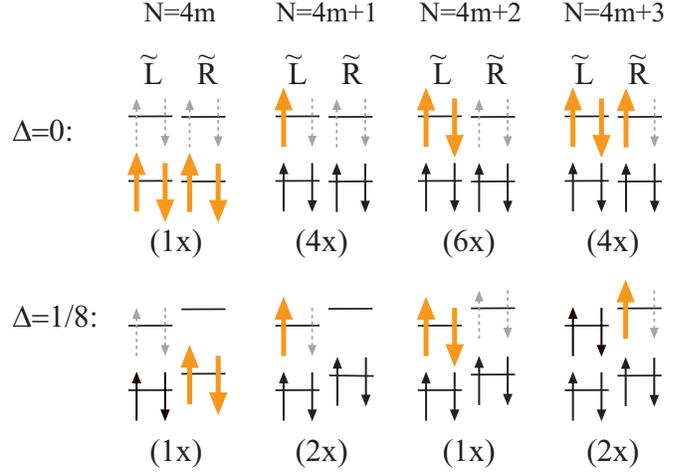}

\caption{\label{cap:GroundStAndCNNp1}Ground states of the SWNT for different
numbers of electrons. For each charge state we show one representative
ground state, on top for aligned bands and on the bottom for mismatched
bands. In brackets the corresponding degeneracy is given. Full black and orange arrows
represent occupied states. The bold (orange) arrows indicate electrons
which can contribute to transitions $N\rightarrow N-1$ in the low
bias regime. Incoming electrons for the transitions $N\rightarrow N+1$
can be accommodated by the states represented by the dashed arrows.
So the number of bold and dashed arrows is equal to $C_{N,N-1}$ and
$C_{N,N+1}$ respectively.}
\end{figure}
We will refer to this kind of master equations as "commonly used master
equations" (CMEs). In the low bias regime the occupation probabilities
$\rho_{\vec{N}\vec{N}}^{I,E_{M}^{0}}$ of the groundstates containing
$M$ particles will all be the same in the stationary solution. Therefore,
we introduce the probability for finding the system in charge state
$M$ by \[
P_{M}=d_{M}\rho_{\vec{N}\vec{N}}^{I,E_{M}^{0}}(t),\]
 where $d_{M}$ is the degeneracy of the corresponding ground states.
Using (\ref{eq:generalized_Mequ}), the general expression for the master equation, we find the following CME for $P_{M}$,
\begin{equation}
\frac{\dot{P}_{M}}{d_{M}}=-R_{\vec{N}\vec{N}\vec{N}\vec{N}}^{E_{M}^0}\frac{P_{M}}{d_{M}}+\sum_{\vec{N}'}R_{\vec{N}\vec{N}\vec{N}'\vec{N}'}^{E_{M}^0E_{M'}^0}\frac{P_{M'}}{d_{M'}}.
\end{equation}
With equations (\ref{eq:Blochredf_1}) and (\ref{eq:Blochredf_2}) for the Redfield tensors we obtain
\begin{multline}
\dot{P}_{M}=-\sum_{l}\left(\Sigma_{l}^{M\rightarrow M'}-\Sigma_{l}^{M'\rightarrow M}\right)=\\
-2\sum_{l}\sum_{\vec{N}'}\left(\Gamma_{l\,\vec{N}\vec{N}'}^{E_{M}^0E_{M'}^0}P_{M}-
\frac{d_{M}}{d_{M'}}\Gamma_{l\,\vec{N}'\vec{N}}^{E_{M'}^0E_{M}^0}P_{M'}\right)\label{eq:CMEPM}, \end{multline}
where we have defined
\begin{equation}
\sum_{\vec{N}'}\Gamma_{l,\vec{N}\vec{N}'}^{E_{M}^0E_{M'}^0}:=\sum_{\vec{N}'}\Gamma_{l\,\vec{N}\vec{N}'\vec{N}'\vec{N}}^{(+)E_{M}^0E_{M'}^0}.
\end{equation}
Using the relation 
\begin{equation}
\frac{d_{M}}{d_{M'}}\sum_{\vec{N}'}\Gamma_{l\,\vec{N}'\vec{N}}^{E_{M'}^0E_{M}^0}=\sum_{\vec{N}'}\Gamma_{l\,\vec{N}\vec{N}'}^{E_{M'}^0E_{M}^0}
\end{equation} the stationary solution of (\ref{eq:CMEPM}) is easily obtained,
\begin{equation}
P_{M}=\frac{\sum_{\vec{N}'}\Gamma_{l\,\vec{N}\vec{N}'}^{E_{M'}^0E_{M}^0}}
           {\Gamma_{tot}},
\end{equation}
with 
\begin{equation}
\Gamma_{tot}=\sum_{l}\sum_{\vec{N}'}\left(\Gamma_{l\,\vec{N}\vec{N}'}^{E_{N}^0E_{N+1}^0}+
                                          \Gamma_{l\,\vec{N}\vec{N}'}^{E_{N+1}^0E_{N}^0}\right).
\end{equation}
% Then the CME for $P_{M}$ is of the form \begin{equation}
% \dot{P}_{M}=-\sum_{l}\left(\Gamma_{l}^{M\rightarrow M'}-\Gamma_{l}^{M'\rightarrow M}\right).\label{eq:MEQPN}\end{equation}
% Here $M'=N+1$ for $M=N$ and $M'=N$ for $M=N+1.$ Equation (\ref{eq:MEQPN})
% is easily solved and we get the stationary solutions \[
% P_{M}=\frac{\sum_{l}\Gamma_{l}^{M'\rightarrow M}}{\Gamma_{tot}},\]
% with $\Gamma_{tot}=\sum_{l}\left(\Gamma_{l}^{N\rightarrow N+1}+\Gamma_{l}^{N+1\rightarrow N}\right).$
The current is obtained from (\ref{eq:CurrentGeneral}). Evaluated
at the source for example it yields,\begin{multline}
I_{N,N+1}=\\
                               \frac{2e\sum_{\vec{N}'}\left(
                               \Gamma_{s\,\vec{N}\vec{N}'}^{E_{N}^0E_{N+1}^0}
                               \Gamma_{d\,\vec{N}\vec{N}'}^{E_{N+1}^0E_{N}^0}-
                               \Gamma_{s\,\vec{N}\vec{N}'}^{E_{N+1}^0E_{N}^0}
                               \Gamma_{d\,\vec{N}\vec{N}'}^{E_{N}^0E_{N+1}^0}\right)}
                              {\Gamma_{tot}}.\label{eq:currentlowbias}\end{multline}
The rates $\Gamma_{l,\vec{N}\vec{N}'}^{E_{M}^0E_{M'}^0}$ are obtained by using equations (\ref{eq:gammaNNp1spec})
and (\ref{eq:gammaNNm1spec}),
\begin{multline*}
\Gamma_{l\,\vec{N}\vec{N}'}^{E_{N}^0E_{N+1}^0}=\\
\frac{\pi}{\hbar}\Phi_{l}\rho_{l}f(\varepsilon_{l})
\sum_{r\sigma}\left(\psi_{r\sigma l}\right)_{\vec{N}\vec{N}'}^{E_{N}^{0}\,E_{N+1}^{0}}
\left(\psi_{r\sigma l}^{\dagger}\right)_{\vec{N}'\vec{N}}^{E_{N+1}^{0}\, E_{N}^{0}},\end{multline*}
 and \begin{multline*}
\Gamma_{l\,\vec{N}\vec{N}'}^{E_{N+1}^0E_{N}^0}=\\
\frac{\pi}{\hbar}\Phi_{l}\rho_{l}\left(1-f(\varepsilon_{l})\right)
\sum_{r\sigma}\left(\psi_{r\sigma l}^{\dagger}\right)_{\vec{N}\vec{N}'}^{E_{N+1}^{0}\, E_{N}^{0}}\left(\psi_{r\sigma l}\right)_{\vec{N}'\vec{N}}^{E_{N}^{0}\, E_{N+1}^{0}}.\end{multline*}
Here we have assumed that $\Phi_{l}$ and $\rho_{l}$ are constant
in the relevant energy range. Furthermore we have defined $\varepsilon_{l}=eV_{l}-\Delta E$
and $\Delta E=E_{N}^{0}-E_{N+1}^{0}.$ From (\ref{eq:Psinm}) we obtain (remember that $\psi_{\tilde{r}\sigma l}:=\psi_{\tilde{r}\sigma K_{0}}(x_{l})/A(x_{l})$),
\[
\sum_{\tilde{r}\sigma}\sum_{\vec{N}'}\left(\psi_{r\sigma l}^{\dagger}\right)_{\vec{N}\vec{N}'}^{E_{M}^{0}\, E_{M'}^{0}}\left(\psi_{r\sigma l}\right)_{\vec{N}'\vec{N}}^{E_{M}^{0}\, E_{M'}^{0}}=\frac{1}{2L}C_{M,M'},\]
where $C_{M,M'}$ is the number of ground states with $M'$ particles
that differ from a given band filling vector $\vec{N}$ for one of the ground state with $M$
particles only by a unit vector (see Fig. \ref{cap:GroundStAndCNNp1}).
Therefore we get \begin{equation}
2\sum_{\vec{N}'}\Gamma_{l\,\vec{N}\vec{N}'}^{E_{N}^0E_{N+1}^0}=\gamma_{l}f(\varepsilon_{l})C_{N,N+1}\label{eq:GNNp1lowbias}\end{equation}
as well as \begin{equation}
2\sum_{\vec{N}'}\Gamma_{l\,\vec{N}\vec{N}'}^{E_{N+1}^0E_{N}^0}=\gamma_{l}\left(1-f(\varepsilon_{l})\right)C_{N+1,N},\label{eq:GNp1Nlowbias}\end{equation}
with $\gamma_{l}=\frac{\pi}{L\hbar}\Phi_{l}\rho_{l}.$ Inserting the
rates (\ref{eq:GNNp1lowbias}) and (\ref{eq:GNp1Nlowbias}) into
expression (\ref{eq:currentlowbias}) for the current results in
\begin{equation}
I_{N,N+1}=e\frac{C_{N,N+1}C_{N+1,N}\gamma_{s}\gamma_{d}\left[f(\varepsilon_{s})-f(\varepsilon_{d})\right]}{\sum_{l}\gamma_{l}\left[f(\varepsilon_{l})C_{N,N+1}+(1-f(\varepsilon_{l}))C_{N+1,N}\right]}.\label{eq:Ilowbias}
\end{equation}

\subsubsection{Linear conductance}

In the regime $\left|eV\right|\ll kT\ll\varepsilon_{0}$ we can further
simplify (\ref{eq:Ilowbias}) by linearising $I_{NN+1}$ in the
bias voltage. We choose $-eV_{s}=eV_{d}=:e\frac{V_{b}}{2}$ and obtain\begin{multline*}
I_{N,N+1}=\\
\frac{e^{2}\beta C_{N,N+1}C_{N+1,N}\gamma_{s}\gamma_{d}}{\sum_{l}\gamma_{l}\left(f_{0}C_{N,N+1}+(1-f_{0})C_{N+1,N}\right)}\frac{e^{-\Delta E}}{\left(e^{-\Delta E}+1\right)^{2}}V_{b},\end{multline*}
with $f_{0}=f(-\Delta E).$ Unlike one could expect, the maxima of
the conductance $G_{N,N+1}=I_{N,N+1}/V_{b}$ are not at $\Delta E=0,$
but at \[
\Delta E_{\mathrm{max}}=\frac{1}{2\beta}ln\frac{C_{N+1,N}}{C_{N,N+1}},\]
which is only zero for $C_{N+1,N}=C_{N,N+1}.$ The height of the conductance
peaks is\begin{multline}
G_{N,N+1}^{\mathrm{max}}=\\
\frac{\gamma_{s}\gamma_{d}C_{N,N+1}C_{N+1,N}}{\left(\gamma_{s}+\gamma_{d}\right)\left(C_{N,N+1}+C_{N+1,N}+2\sqrt{C_{N,N+1}C_{N+1,N}}\right)}e^{2}\beta.\label{eq:Gmax}\end{multline}
We still have to determine the values of $C_{N,N+1}$ and $C_{N+1,N},$
which depend on the mismatch between the $\tilde{R}$ and the $\tilde{L}$
band. If both bands are aligned $(\Delta=0,\,1/2)$ one finds from
Fig. \ref{cap:GroundStAndCNNp1}, $C_{N,N+1}=4,3,2,1$ and $C_{N+1,N}=1,2,3,4$
for $N=4m,\,4m+1,\,4m+2,\,4m+3.$ Then the conductance $G_{N,N+1}$
shows fourfold electron periodicity with two equally high central
peaks for $N=4m+1,\,4m+2$ and two smaller ones for $N=4m,\,4m+3$
(cf. Fig. \ref{cap:ConductanceDelta0} a)). The relative height between
central and outer peaks is $G_{4m+1,4m+2}^{\mathrm{max}}/G_{4m,4m+1}^{\mathrm{max}}=27/(10+4\sqrt{6})\approx1.36.$
Note that this ratio is \textit{independent} of a possible asymmetry $\gamma_{s}\neq\gamma_{d}$
in the lead contacts. In addition, the conductance is symmetric under
an exchange of the sign of the bias voltage.

In the case of an energy mismatch between the $\tilde{R}$ and the
$\tilde{L}$ band exceeding well the thermal energy, the degeneracy
of the ground states is either $1$ or $2$ and we get $C_{N,N+1}=2,1,2,1$
and $C_{N,N+1}=1,2,1,2$ for $N=4m,\,4m+1,\,4m+2,\,4m+3.$ Therefore,
according to (\ref{eq:Gmax}), all the conductance peaks have the
same height as depicted in Fig. \ref{cap:ConductanceDelta0.4} a)
.

\subsubsection{Low bias regime well outside of the Coulomb diamonds}

\begin{figure}
\includegraphics[%
  clip,
  width=1.0\columnwidth]{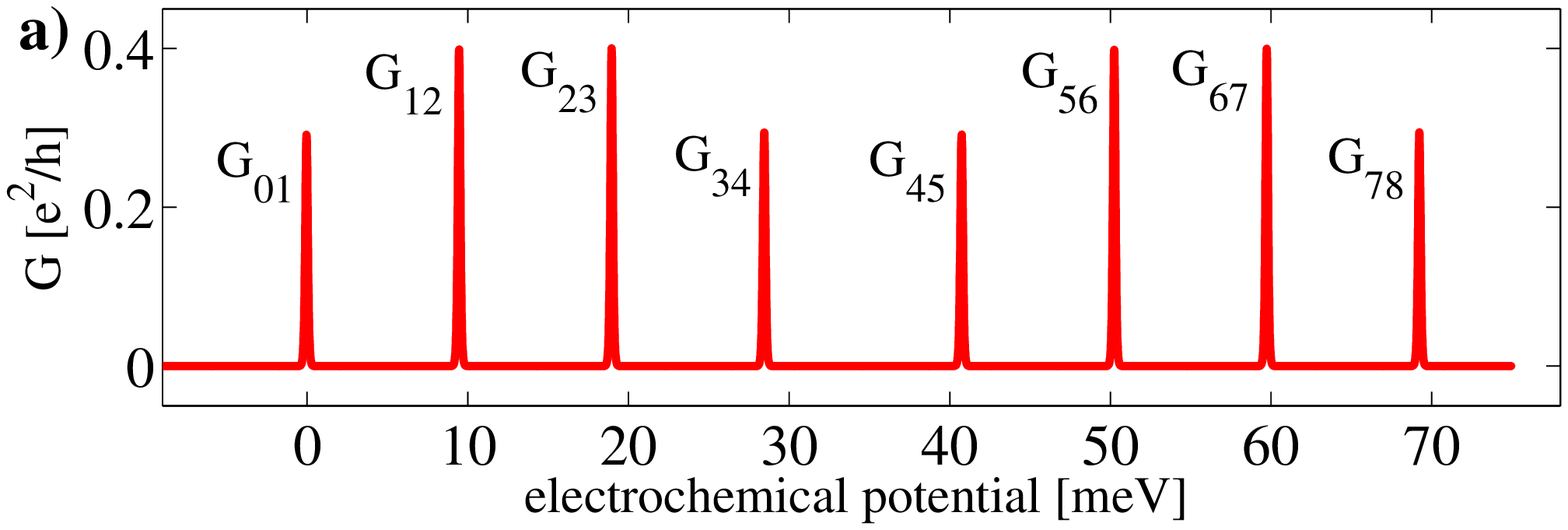}

\includegraphics[%
  clip,
  width=1.0\columnwidth]{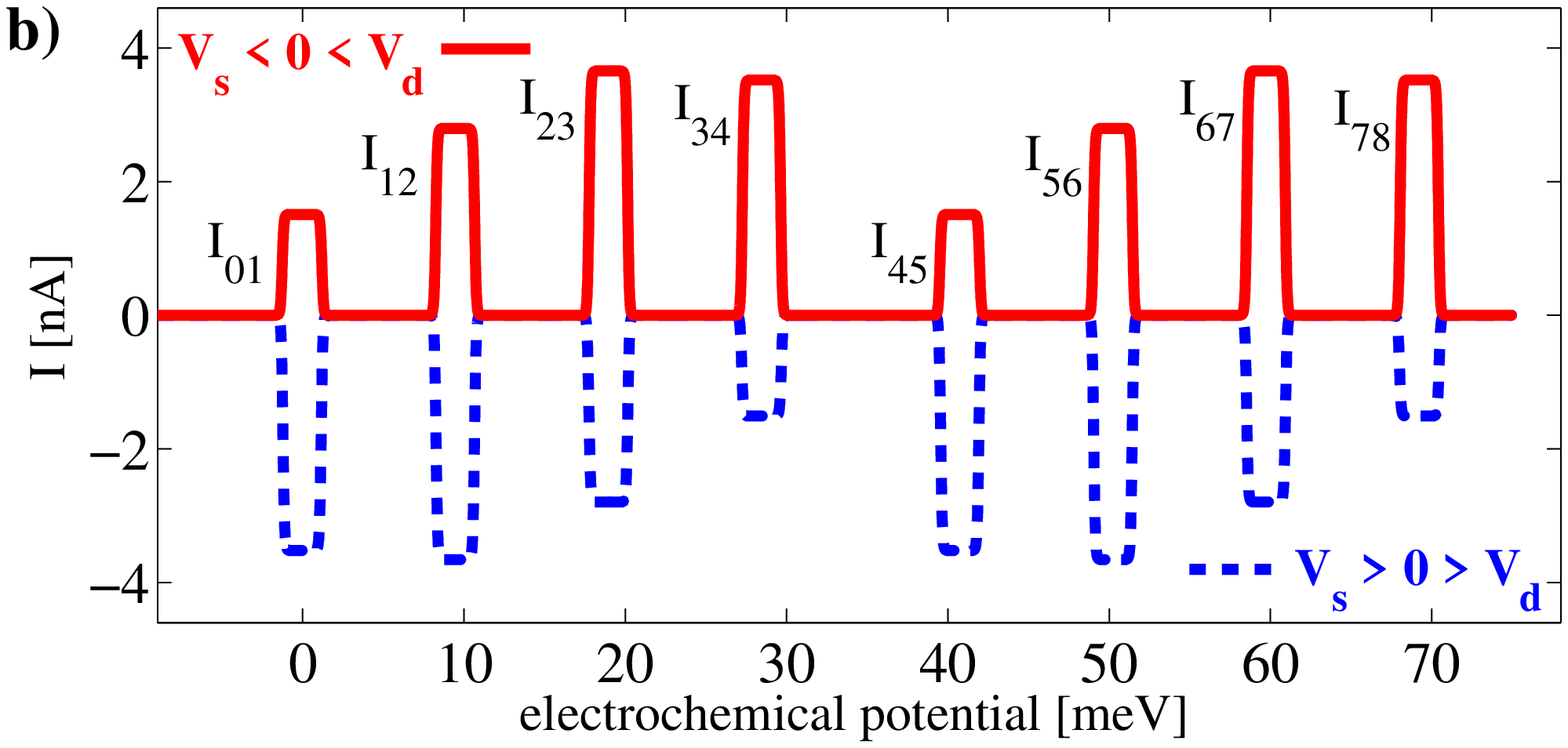}

\caption{\label{cap:ConductanceDelta0}Gate traces for $\Delta=0$, i.e. for
aligned $\tilde{R}$ and $\tilde{L}$ bands, at low bias voltage.
Parameters are $E_{c}=9.5$ meV, $k_{B}T=0.10$ meV, $\varepsilon_{0}=2.9$ meV
and the chosen asymmetry is $\gamma_{s}=5\gamma_{d}=4.9\cdot10^{10}\textrm{s}^{-1}$.
a) Conductance in the linear regime $eV_{b}\ll k_{B}T\ll\varepsilon_{0}.$
Despite asymmetric contacts, we find the repeating pattern of two
small outer peaks and two large central peaks. b) Current in the regime
$k_{B}T\ll\left|eV_{l}\right|\ll\varepsilon_{0}.$ Asymmetry effects
appear. The fourfold periodicity is retained. The upper and lower
pattern correspond to opposite values of the bias voltage.}
\end{figure}

\begin{figure}
\includegraphics[%
  width=1.0\columnwidth]{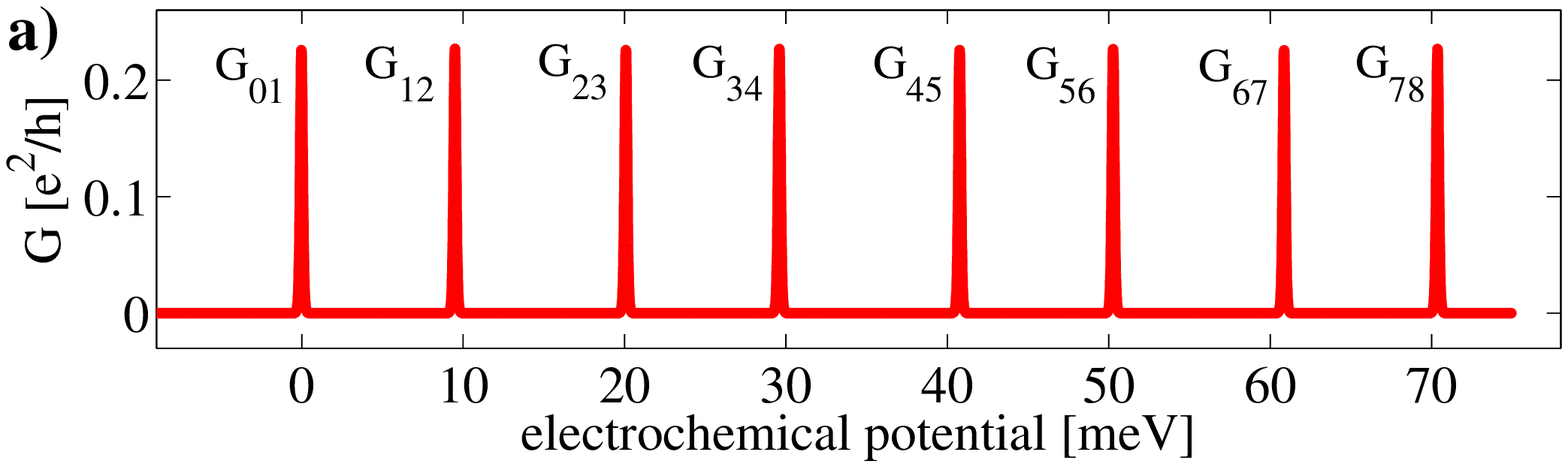}

\includegraphics[%
  clip,
  width=1.0\columnwidth]{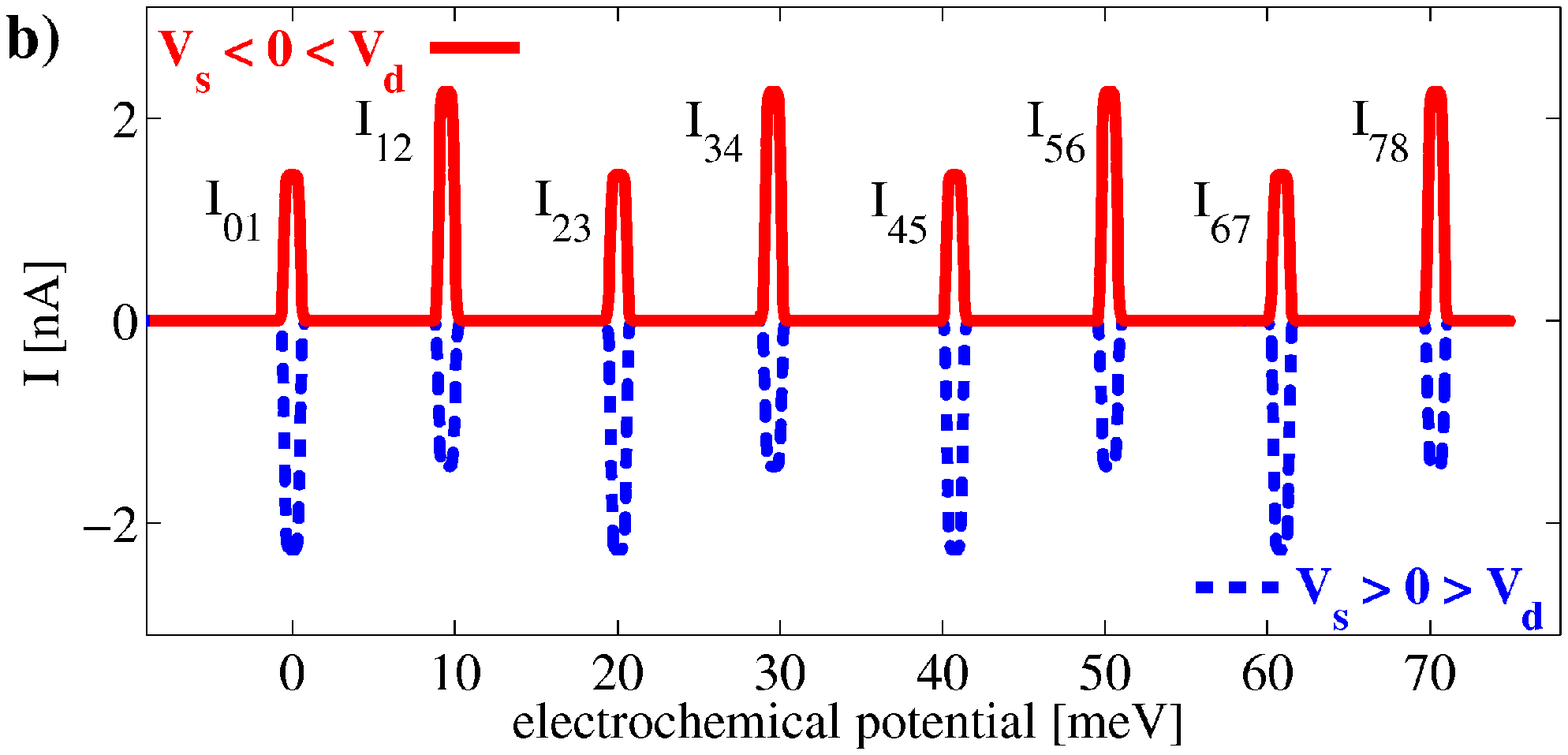}

\caption{\label{cap:ConductanceDelta0.4}Gate traces for $\Delta=0.2$, i.e.
for mismatched $\tilde{R}$ and $\tilde{L}$ bands, at low bias voltage.
Other parameters are $E_{c}=9.5$ meV, $k_{B}T=0.10$ meV, $\varepsilon_{0}=2.9$ meV
and the chosen asymmetry is $\gamma_{s}=5\gamma_{d}=4.9\cdot10^{10}\textrm{s}^{-1}$.
a) Conductance in the linear regime $eV_{b}\ll k_{B}T\ll\varepsilon_{0}.$
The conductance peaks are all of the same height. b) Current in the
regime $k_{B}T\ll\left|eV_{l}\right|\ll\varepsilon_{0}.$ The contact
asymmetry leads to alternating large and small current maxima.}
\end{figure}

Now we examine the regime where still only ground states are occupied
but where we are well outside the region of Coulomb blockade, i.e. $\varepsilon_{0}\gg\left|eV_{l}\pm\Delta E\right|\gg kT.$
Then we have $\left|f(\varepsilon_{s})-f(\varepsilon_{d})\right|=1.$ If
e.g. $eV_{s}-\Delta E<0$ and $eV_{d}-\Delta E>0,$ such that electrons tunnel
in from the source and tunnel out at the drain, we have $f(\varepsilon_{s})=1$ and
$f(\varepsilon_{d})=0$ such that (\ref{eq:Ilowbias}) becomes \begin{equation}
I_{N,N+1}=e\frac{C_{N,N+1}C_{N+1,N}\gamma_{s}\gamma_{d}}{\gamma_{s}C_{N,N+1}+\gamma_{d}C_{N+1,N}}.\label{eq:Ilhighlowbias}\end{equation}
The height of the plateaus in the current of Figures \ref{cap:ConductanceDelta0}
b) and \ref{cap:ConductanceDelta0.4} b) are described by (\ref{eq:Ilhighlowbias}).
If the $\tilde{R}$ band is aligned with the $\tilde{L}$ band, we
still find a fourfold electron periodicity. But only for $\gamma_{s}=\gamma_{d}$
the pattern with two central peaks and two smaller outer peaks is
preserved. The corresponding ratio of the heights is $3/2.$ If $\gamma_{s}\neq\gamma_{d}$
this latter symmetry is lost.

For mismatched bands and $\gamma_{s}=\gamma_{d}$ we find like for
the conductance peaks that all current maxima are of the same size.
In the case of asymmetric tunnelling contacts, $\gamma_{s}\neq\gamma_{d},$
a pattern of alternating small and large peaks is found.

If we invert the sign of the bias voltage, the current is obtained
by flipping its direction and exchanging $\gamma_{s}$ with $\gamma_{d}$
in (\ref{eq:Ilhighlowbias}). Then, if $\gamma_{s}\neq\gamma_{d},$
the current does not only change its sign but also changes its magnitude
because of $C_{N,N+1}\neq C_{N+1,N}$.

\subsection{\label{sub:High-bias-regime}High bias regime}

\begin{figure}
\centering
\begin{minipage}{1.0\columnwidth}
\includegraphics[%
  clip,
  width=1.0\columnwidth]{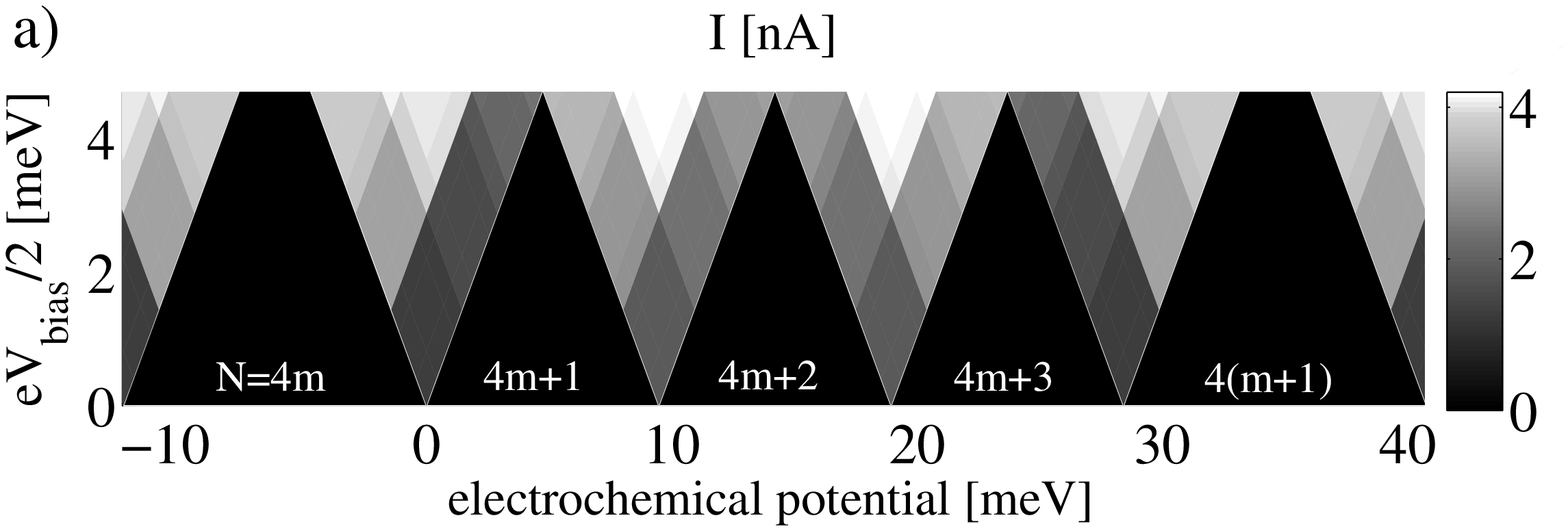}
\end{minipage}
\\[5pt]
\begin{minipage}{1.0\columnwidth}
\includegraphics[%
  clip,
  width=1.0\columnwidth]{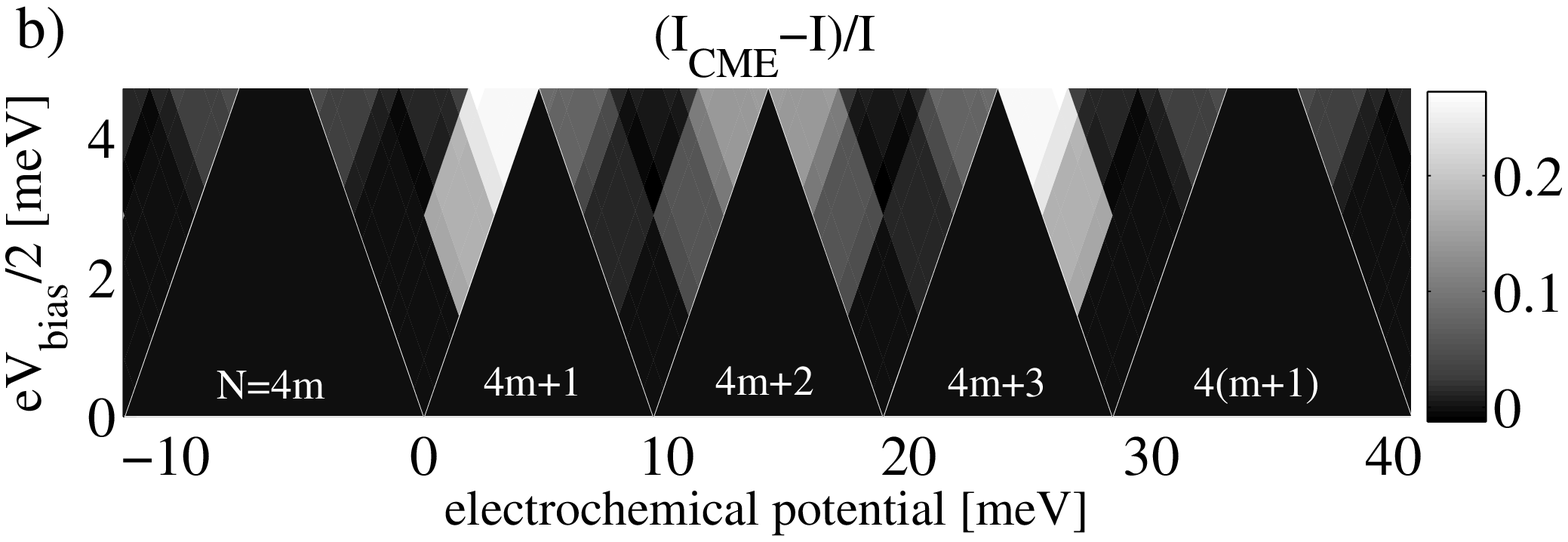}
\end{minipage}

\caption{\label{cap:IDelta0}a) Current in a bias voltage - electrochemical
potential plane for the symmetric contacts case. b) Difference plot
of the current with and without coherences. Here $k_{B}T=0.01\,\mathrm{meV}$
and the interaction parameter is $g\approx0.21$. Other parameters are as for Fig. \ref{cap:ConductanceDelta0}.}
\end{figure}

\begin{figure}
\includegraphics[%
  width=1.0\columnwidth]{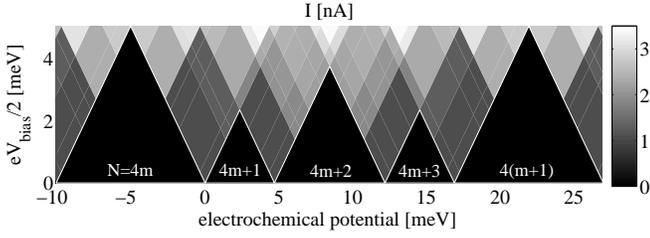}

\caption{\label{cap:IDelta0.4} Current as a function of the bias voltage
and the electrochemical potential in the SWNT for symmetric contacts.
The parameters here are chosen to fit the positions of the Coulomb
diamonds of sample A in \cite{Sapmaz2005}, i.e. $E_{c}=4.7$ meV,
$2\Delta\varepsilon_{0}=2.8$ meV, $\varepsilon_{0}=8.2$ meV. }
\end{figure}

In the bias regime $eV>\varepsilon_{0}$ not only the ground states will
contribute to transport but also states with bosonic excitations and
band filling configurations $\vec{N}$ different from the ground state
configurations. We refer to the latter type of excitations as fermionic
excitations. Since the number of relevant states increases rapidly
with increasing bias voltage, an analytical treatment is not possible
any more and we have to resort to numerical methods in order to calculate
the stationary solution of the master equation (\ref{eq:generalized_Mequ})
and the respective current. From (\ref{eq:exclines}) we know that
at low temperatures the current only changes considerably near the
excitation lines given therein. Therefore we can reduce drastically
the number of $(eV,\ \mu_{g})$ points for which we actually perform
the numerical calculations, saving computing time. In Figs. \ref{cap:IDelta0}
a) and \ref{cap:IDelta0.4}, the current as a function of the applied
bias voltage and the electrochemical potential in the dot is depicted.
The chosen parameters for $E_{c},\,\varepsilon_{0}$ and $\Delta$
are the ones we have obtained for fitting the data of sample C and
sample A of \cite{Sapmaz2005}, respectively. Hence Figs. \ref{cap:IDelta0} and \ref{cap:IDelta0.4}
show the current for SWNTs with the $\tilde{R}$ and $\tilde{L}$
band being aligned ($\Delta=0$) and mismatched ($\Delta\approx0.17$).
In both cases a symmetric coupling to the leads, i.e. $\gamma_{s}=\gamma_{d},$
is assumed. In the transport calculations the lowest lying $c+$ excitations
are taken into account using $g=0.21$ for the Luttinger parameter (cf. equation (\ref{eq:g}) for the definition of $g$).

In addition we have also determined the current using the CME, hence
ignoring any coherences in the RDM. For the current corresponding
to Fig. \ref{cap:IDelta0}, the quantitative difference between the
calculations with and without coherences is considerable in the region
of intermediate bias voltage as we show in Fig. \ref{cap:IDelta0}
b). On the other hand the deviation of the CME result from the calculation
including coherences is by far less pronounced for the parameter choice
of Fig. \ref{cap:IDelta0.4}. The crucial point here is that the charging
energy $E_{c}$ is smaller than $\varepsilon_{0}$, the level spacing
of the neutral system. Then the subsequent considerations are not
strictly valid. Now we explain why coherences can't be generally ignored
if considering interacting electrons in a SWNT.

\subsubsection{Why and when are coherences needed?\label{par:Why-and-when}}

As in the low bias regime we assume to have unpolarized leads, i.e.
we use condition (\ref{eq:NoPol}), such that we can ignore coherences
between states with different fermionic configurations $\vec{N}$.
Unlike in the low bias regime, we are not only left with diagonal elements
of the RDM but there still might be coherences between degenerate
states which have the same $\vec{N}$ but different bosonic excitations
$\vec{m}.$ For the importance of these kind of coherences it is illuminating
to discuss our system without electron-electron interactions, i.e.
for the moment let us assume that an eigenbasis of $H_{\odot}$ is
given by the Slater determinants of the single electron states $\left|\varphi_{\tilde{r}\kappa}^{OBC}\right\rangle \left|\sigma\right\rangle .$
Furthermore, we concentrate without loss of generality on the case
$\Delta=0$, and we assume that the charging energy $E_{c}$
exceeds $\varepsilon_{0}$. Each of the Slater determinants can be
denoted by the occupation $\vec{n}$ of the single electron states.
In the case condition (\ref{eq:NoPol}) holds, it is again easy to
show that coherences vanish in the stationary solution of the master
equation (\ref{eq:generalized_Mequ}), if the RDM is expressed in
the $\left|\vec{n}\right\rangle $ basis. But of course we still could
use the states $\left|\vec{N},\vec{m}\right\rangle $ from (\ref{eq:eigenstates})
as eigenbasis, now with four neutral modes $c+,c-,s+,s-.$ In the
$\left|\vec{N},\vec{m}\right\rangle $ basis it is crucial to include
the off diagonal elements in order to get the right stationary solution
as we show in the following example. 

\begin{figure}
\includegraphics[%
  width=1.0\columnwidth]{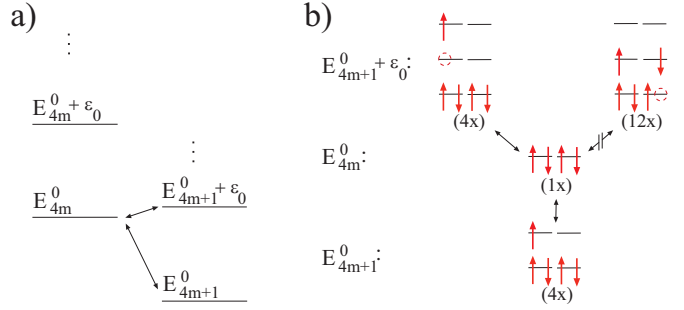}
\caption{\label{cap:The-occupied-}a) Scheme of relevant energy
levels, if transitions from the ground state with $N=4m$ particles
to the first excited state with $N=4m+1$ particles are energetically
allowed, while no transitions from the $N=4m+1$ ground states to
excited states with $N=4m$ electrons are possible. b) Possible transitions between  the $N=4m$ and $N=4m+1$ electron states of the
noninteracting system, which are energetically allowed in situation a). The degeneracy of the
eigenstates is given in brackets.}
\end{figure}

We adjust the voltages such that
only transitions are possible from the ground state with $4m$ particles
and energy $E_{4m}^{0}$ to ground and first excited states with $4m+1$
particles and energies $E_{4m+1}^{0}$ and $E_{4m+1}^{1}=E_{4m+1}^{0}+\varepsilon_{0}$,
respectively. From the $4m+1$ ground states only transitions to the
$4m$ ground state shall be allowed as depicted in Fig.
\ref{cap:The-occupied-} a). Note that this situation is only stable,
because we have made the choice $E_{c}>\varepsilon_{0}$. The master
equation expressed in the $\left|\vec{n}\right\rangle $ basis reveals
that only four of the 16 states with the lowest particle hole excitation
are indeed occupied in the stationary limit (cf. side b) of Fig.
\ref{cap:The-occupied-}, because not all of the corresponding energetically
allowed transitions from the $N=4m$ ground state can be mediated by one-electron tunnelling processes.
Whereas in the $\left|\vec{N},\vec{m}\right\rangle $ basis, all 16
states with the energetically lowest bosonic excitations are equally
populated. Since the degenerate states of the two bases are connected
by a unitary transformation, the same must be true for the corresponding
matrix representations of $\rho_{\odot}^{I,E_{4m+1}^{1}}.$ 
From the representation of the RDM in the $\left|\vec{n}\right\rangle $
basis, we know that the rank of $\rho_{\odot}^{I,E_{4m+1}^{1}}$ must be equal to $4$.
Because an unitary transformation does not change the rank of a matrix, the stationary
solution in the $\left|\vec{N},\vec{m}\right\rangle $ basis can maximally have 4 linearly independent columns. Since all diagonal elements are nonvanishing this is only possible if there are also nonvanishing coherences.

Switching on the electron - electron interactions, the $\left|\vec{n}\right\rangle $
states are no longer an eigenbasis of the SWNT Hamiltonian and hence
we must work in the $\left|\vec{N},\vec{m}\right\rangle $ basis.
But then, as we know from the discussion above, the coherences are
expected to be of importance.

\subsubsection{Negative differential conductance}

\begin{figure}
\includegraphics[%
  clip,
  width=1.0\columnwidth]{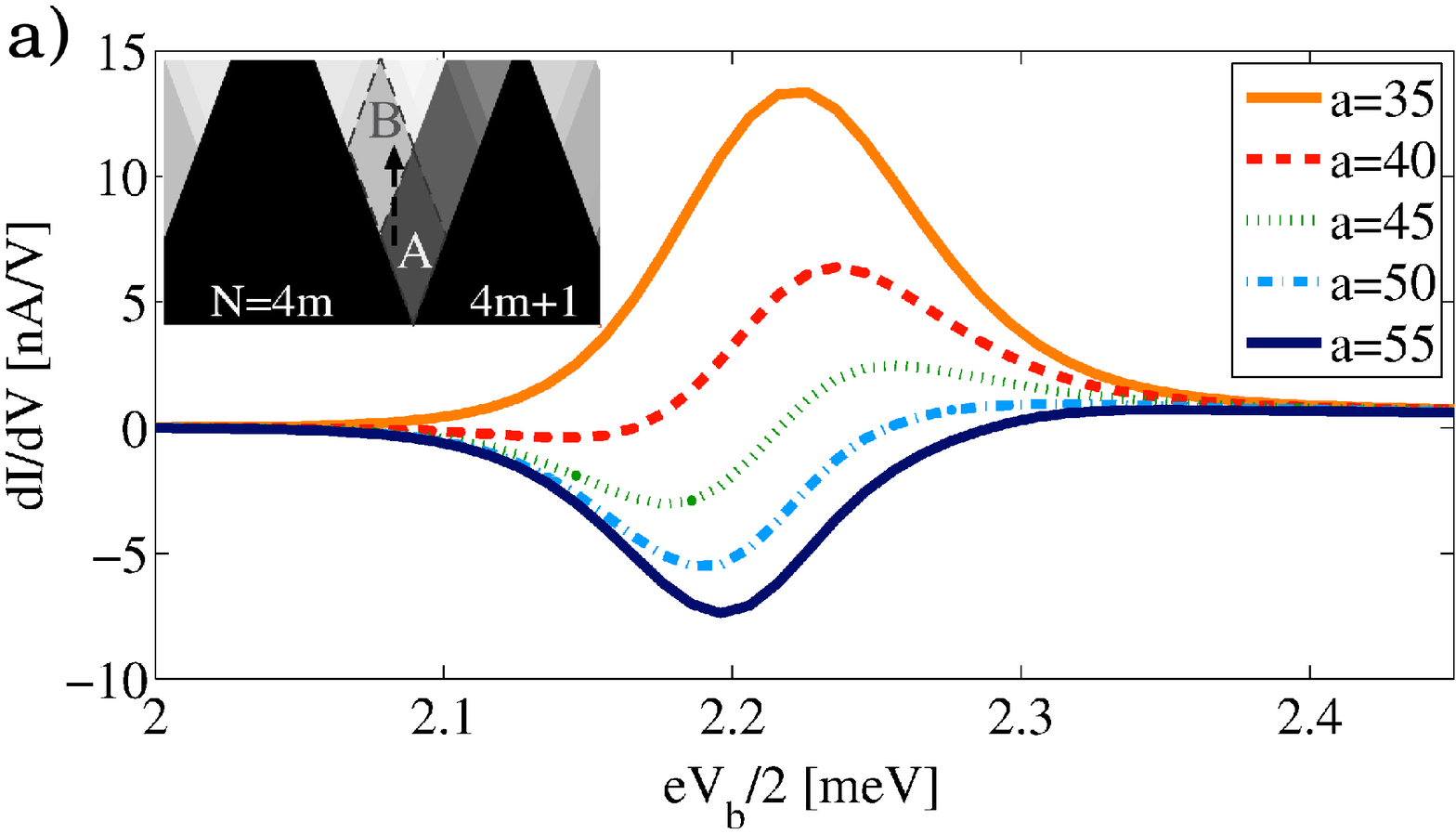}

\includegraphics[%
  clip,
  width=1.0\columnwidth]{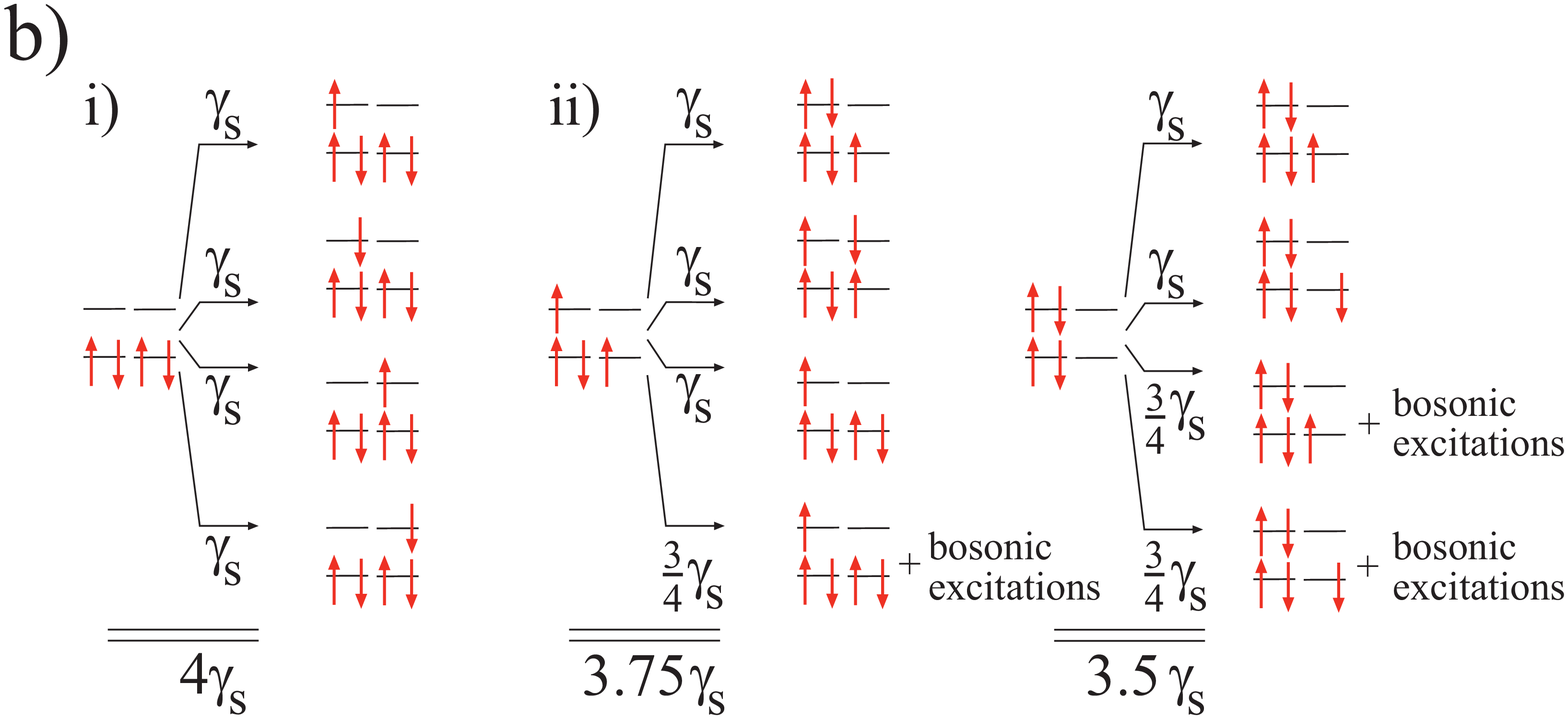}

\caption{\label{cap:NDC}a) Differential conductance as a function of the
bias voltage for different asymmetries $a=\gamma_{d}/\gamma_{s}$
of the coupling to the leads. The trace in the $\mu_{b}$-$V_{b}$
plane across the boundary between regions A and B is indicated in
the inset. Only for asymmetries larger than around $45$ negative
differential conductance occurs. All parameters here are chosen as
for Fig. \ref{cap:IDelta0}, except for $k_{B}T$ a value of $0.026$
meV was used. b) The transitions responsible for NDC. i) In region
A of the inset from a) only transitions between ground states are
possible at low enough temperature. The transition rate from the $N=4m$
ground state to the $N=4m+1$ ground states is given by $4\gamma_{s}$
(see equation (\ref{eq:GNNp1lowbias})). ii) In region B additionally
excited states become occupied. For some of the states with $N=4m$
electrons and fermionic excitations the transition rates to states
with $N=4m+1$ electrons and neutral bosonic excitations is decreased
compared to the ground state rate as a consequence of the larger
energies of $c+$ excitations, which are not available for transport
yet. Notice that we only show the most important types of transitions
from $N=4m$ to $N=4m+1$ that take place in region B. Since we are
considering a nonequilibrium situation other types of transitions
are possible in principle.}
\end{figure}

Spin charge separation and therefore non-Fermi liquid behavior could
also manifest itself in the occurrence of negative differential conductance
(NDC) at certain excitation lines involving transitions to states
with fermionic excitations, as was predicted for a spinful Luttinger
liquid quantum dot \cite{Cavaliere2004} with asymmetric contacts.
We also find this effect for the nonequilibrium treatment of the SWNT
quantum dot. Since in our case only the energy spectrum of the $c+$
mode depends on the interaction and the other three modes have the
same energies as the neutral system, rather large asymmetries are
needed in order to observe NDC. In Fig. \ref{cap:NDC} we show the
current across the first excitation line for transitions from $N=4m+1$
to $N=4m$ in the $\Delta=0$ case. The corresponding trace in the
$\mu_{b}$-$V_{b}$ plane is indicated in the inset of Fig. \ref{cap:NDC}
a). Here the origin of the NDC is that some states with fermionic
excitations have lower transition rates than nonexcited states, since
due to the increased energy of the $c+$ modes less channels are available
for transport. In Fig. \ref{cap:NDC} b) we show some of the excited
states with $N=4m$ electrons which are responsible for the NDC, because
their transition amplitudes to states with $N=4m+1$ electrons are
reduced compared with the one of the $4m$ ground state. Apart from
the asymmetries all other parameters are chosen as for Fig. \ref{cap:IDelta0}.
Only for asymmetries $a=\gamma_{d}/\gamma_{s}$ larger than around
$45$, clear NDC features are seen.%

\section{\label{sec:Conclusions}Conclusions}
In this article we have analyzed the linear and nonlinear current as a function of the gate and bias voltage across metallic SWNT quantum dots. The properties of the SWNT itself were derived from a microscopic model including electron - electron interactions. Exchange and related effects, which become relevant for small diameter SWNTs, have not been taken into account. The energy spectrum of a metallic SWNT, which can be obtained with the help of bosonization, turned out to be highly degenerate as a consequence of both fermionic and bosonic excitations. In the linear bias regime, the degeneracy of the groundstates leads to a characteristic pattern of conductance peaks depending on whether the two branches of the dispersion relation are aligned or not. Leaving the linear regime, asymmetry effects become relevant. Thus measurements of the current at low bias voltages, in the linear and nonlinear regime, in principle allow the separate determination of the source and drain tunnelling resistances as a function of the gate voltage. 
At higher bias voltages also excited states become relevant. The correct calculation of the nonequilibrium dynamics of the system then requires the inclusion of coherences in the reduced density matrix between degenerate states with bosonic excitations. At intermediate bias voltages there is a considerable deviation between the transport calculations with and without coherences. We emphasize that for a noninteracting system with unpolarized leads, coherences do not have to be considered if expressing the reduced density matrix in terms of Slater determinants, formed by the one electron wave functions of the noninteracting system.
Another consequence of the electron correlations is the formation of a non-oscillatory spatial dependence of the tunnelling amplitudes along the nanotube axis. For transitions between states with energetically low  excitations we find a strong suppression of the tunnelling amplitudes near the SWNT ends.
Furthermore we have addressed the influence of the tunnelling contacts on the transport. We have shown that extended contacts described as 3D Fermi gas do not lead to a polarization of the contacts with respect to the two branches of the dispersion relation. We think that a further investigation of this point for other types of contacts is worthwhile.

\centerline{{*}{*}{*}} Useful discussions with S. Sapmaz and support
by the DFG under the program GRK 638 are acknowledged.

\appendix

\section{\label{sec:Description-of-the}Description of the tunnelling contacts}

In this appendix we discuss the dependence of the rates $\Gamma_{l,k'mnk}^{(\alpha)E_{M}\, E_{M'}}$
on the properties of the tunnelling contacts. In Section \ref{sub:Current-trough-a}
we have introduced the parameters $\Phi_{lrr'}(\varepsilon)$ with equation
(\ref{eq:Phirr}). Here we show the calculation of these parameters
assuming a 3D electron gas in the leads. Furthermore, we require that
the tunnelling region extends over \textit{several sites} of the SWNT lattice.
Starting from (\ref{eq:gamma_lEnEnp1}) together with (\ref{eq:CalF_sigma_l}) and  (\ref{eq:Psisigmaexp}) 
we can characterize the part of $\Gamma_{l,k'mnk}^{(\alpha)E_{N}\, E'_{N+1}}$ that depends on the tunnelling contact $l$ by the following expression,
\begin{multline}
\mathcal{T}_{l,nkk'm}(\varepsilon):=\sum_{FF'}\mathrm{sgn}(FF')\int d^{3}r\int d^{3}r'T_{l}^{*}(\vec{r})T_{l}(\vec{r}')\times\\
\sum_{\vec{q}|_{\varepsilon}}\phi_{l\vec{q}}^{*}(\vec{r})\phi_{l\vec{q}}(\vec{r}')\sum_{\tilde{r}\tilde{r}'\sigma}\tilde{\varphi}_{\mathrm{sgn}(F)\tilde{r}\, F}(\vec{r})\tilde{\varphi}_{\mathrm{sgn}(F')\tilde{r}'\, F'}^{*}(\vec{r}')\times\\
\left(\psi_{\tilde{r}F\sigma}(x)\right)_{nk}\left(\psi_{\tilde{r}'F'\sigma}^{\dagger}(x')\right)_{k'm}.\label{eq:Tauvoneps}\end{multline}
The wave functions in the leads are denoted $\phi_{l\vec{q}}$ and
the sum $\sum_{\vec{q}|_{\varepsilon}}$ extends over all $\vec{q}$
values that correspond to the energy $\varepsilon.$ At low enough
energies, oscillations of the product $\left(\psi_{\tilde{r}F\sigma}(x)\right)_{nk}\left(\psi_{\tilde{r}'F'\sigma}^{\dagger}(y)\right)_{k'm}$
can be ignored along the length of the tunnelling interfaces. Using
(\ref{eq:Psinm}) we thus can rewrite (\ref{eq:Tauvoneps}) as \begin{multline}
\mathcal{T}_{l,nkk'm}(\varepsilon)=L\sum_{FF'}\mathrm{sgn}(FF')\nu_{lFF'}(\Delta)\times\\
\int d^{3}r\int d^{3}r'T_{l}^{*}(\vec{r})T_{l}(\vec{r}')A(x)A(x')\sum_{\vec{q}|_{\varepsilon}}\phi_{l\vec{q}}^{*}(\vec{r})\phi_{l\vec{q}}(r')\times\\
\sum_{\tilde{r}\tilde{r}'\sigma}\varphi_{\mathrm{sgn}(F)\tilde{r}\, F}(\vec{r})\varphi_{\mathrm{sgn}(F')\tilde{r}'\, F'}^{*}(\vec{r}')\left(\psi_{\tilde{r}\sigma l}\right)_{nk}\left(\psi_{\tilde{r}'\sigma l}^{\dagger}\right)_{k'm}.\end{multline}
Here the factor $\nu_{lFF'}(\Delta)=e^{i\delta_{l,d}\pi\mathrm{sgn}(F-F')\Delta}$
takes into account a possible phase shift due to the band mismatch
$\Delta$ and we have used the notation $\psi_{\tilde{r}\sigma l}=\psi_{\tilde{r}K_{0}\sigma}(x_{l})/A(x_{l}),$
where $x_{l}=0,\, L$ for $l=s,\, d.$ Now the parameters $\Phi_{l\tilde{r}\tilde{r}'}(\varepsilon)$
from (\ref{eq:Phirr}) are recovered by the relation\[
\mathcal{T}_{l,nkk'm}(\varepsilon)=\sum_{\tilde{r}\tilde{r}'}\Phi_{l\tilde{r}\tilde{r}'}(\varepsilon)\sum_{\sigma}\left(\psi_{\tilde{r}\sigma l}\right)_{nk}\left(\psi_{\tilde{r}'\sigma l}^{\dagger}\right)_{k'm}.\]
 In the next step we exploit that the Bloch waves $\varphi_{\mathrm{sgn}(F)r\, F}$
from equation (\ref{eq:sublatticewaves}) are only nonvanishing around
the positions of the carbon atoms in the SWNT lattice. On the length
scale of the extension of the $p_{z}$ orbitals all other quantities
in $\Phi_{l\tilde{r}\tilde{r}'}$ are slowly varying. Hence we can
rewrite the integrals over $\vec{x}$ and $\vec{y}$ as a sum over
the positions of the carbon atoms\begin{multline}
\Phi_{l\tilde{r}\tilde{r}'}(\varepsilon)=C\frac{L}{N_{L}}\sum_{FF'}\mathrm{sgn}(FF')\nu_{lFF'}(\Delta)\times\\
\sum_{p,p'}f_{p\mathrm{sgn}(F)r\, F}f_{p'\mathrm{sgn}(F')r'\, F'}^{*}\sum_{\vec{R},\vec{R}'}T_{l}^{*}(\vec{x}_{\vec{R},p})T_{l}(\vec{x}_{\vec{R}',p'})\times\\
A(\vec{R}_{x})A(\vec{R}'_{x})\sum_{\vec{q}|_{\varepsilon}}\phi_{l\vec{q}}^{*}(\vec{x}_{\vec{R},p})\phi_{l\vec{q}}(\vec{x}_{\vec{R}',p'})e^{iF\vec{R}}e^{-iF\vec{R}'},\label{eq:Tvoneps2}\end{multline}
where we have defined $\vec{x}_{\vec{R},p}:=\vec{R}+\vec{\tau}_{p}.$
The constant $C$ results from the integration over the $p_{z}$ orbitals.
For the 3D electron gas the wave functions $\phi_{l\vec{q}}(\vec{x})$
are simply given by plane waves,\[
\phi_{l\vec{q}}(\vec{x})=\frac{1}{\sqrt{V_{l}}}e^{i\vec{q}\vec{x}}.\]
 Therefore we can easily perform the sum over the wave numbers $\vec{q}$
associated with the energy $\varepsilon$,\begin{multline*}
\sum_{\vec{q}|_{\varepsilon}}\phi_{l\vec{q}}^{*}(\vec{x})\phi_{l\vec{q}}(\vec{y})\approx\\
\int_{0}^{2\pi}d\varphi\int_{-1}^{1}d\cos\theta e^{iq|_{\varepsilon}\left|\vec{x}-\vec{y}\right|\cos\vec{\theta}}=\frac{4\pi\sin(q|_{\varepsilon}\left|\vec{x}-\vec{y}\right|)}{q|_{\varepsilon}\left|\vec{x}-\vec{y}\right|}.\end{multline*}
The previous expression is peaked around $\vec{x}=\vec{y}.$ For $q|_{\varepsilon}>1/\left|\vec{x}-\vec{y}\right|$
the correlations between $\vec{x}$ and $\vec{y}$ drop off fast.
It means that if $q|_{\varepsilon}$ is larger than $1/a_{0},$ where
$a_{0}$ is the nearest neighbor distance on the SWNT lattice, correlations
between different carbon atom sites are suppressed such that we arrive
at the following approximation:\[
\sum_{\vec{q}|_{\varepsilon}}\phi_{l\vec{q}}^{*}(\vec{x}_{\vec{R},p})\phi_{l\vec{q}}(\vec{x}_{\vec{R}',p'})\approx4\pi\delta_{\vec{R},\vec{R}'}\delta_{p,p'},\]
which yields \begin{multline}
\Phi_{l\tilde{r}\tilde{r}'}(\varepsilon)=4\pi C\frac{L}{N_{L}}\times\\
\sum_{FF'}sgn(FF')\nu_{lFF'}(\Delta)\sum_{p}f_{p\mathrm{sgn}(F)r\, F}f_{p\mathrm{sgn}(F')r'\, F'}^{*}\times\\
\sum_{\vec{R}}\left|T_{l}(\vec{x}_{\vec{R},p})\right|^{2}e^{i\left(F-F'\right)\vec{R}}A^{2}(\vec{R}_{x}).\label{eq:Tvoneps3}\end{multline}
Since we assume an extended tunnelling region, the fast oscillating
terms with $F=-F'$ are supposed to cancel. Furthermore we assume that both sublattices are equally well coupled to the contacts, such that the sum over $\vec{R}$ in (\ref{eq:Tvoneps3}) should approximately give the same
result for $p=1$ and $p=2.$ Hence we can separate the sum over $p$
from the rest. Since the Bloch waves $\tilde{\varphi}_{\mathrm{sgn}(F)r\, F}$
and $\tilde{\varphi}_{\mathrm{sgn}(F)r'\, F}$ are orthogonal to each
other for $r\neq r'$, the relation \[
\sum_{p}f_{p\mathrm{sgn}(F)r\, F}f_{p\mathrm{sgn}(F)r'\, F}^{*}=\delta_{rr'}\]
must hold, as it can be deduced from the explicit expression for the
Bloch wave, equation (\ref{eq:sublatticewaves}), and we arrive at\begin{multline}
\Phi_{l\tilde{r}\tilde{r}'}(\varepsilon)=\delta_{rr'}4\pi C\frac{L}{N_{L}}\sum_{\vec{R}}\left|T_{l}(\vec{x}_{\vec{R},p})\right|^{2}A^{2}(\vec{R}_{x}).\end{multline}
We should mention that the treatment of the contact geometry here
relies on several assumptions that might not be fulfilled under all
circumstances and so it should be seen as a first estimate. For example
we assume that the transmission functions $T_{l}$ do not depend on
the $\vec{q}$ vector of the incoming wave. But such a dependence
would increase the correlation between different atom sites. Furthermore
$q|_{\varepsilon}$ will not be much larger than $1/a_{0}$ for all kind
of contacts (for gold the Fermi wave number is about $1.5/a_{0}),$
nor can the lead electrons always be described by a 3D electron gas. 

Here we want to emphasize, the tunnelling contacts itself play an important
role for transport through a SWNT quantum dot. Especially it will
be interesting for which kind of contacts the condition \[
\Phi_{l\tilde{r}\tilde{r}'}(\varepsilon)\neq\delta_{\tilde{r}\tilde{r}'}\Phi_{l}(\varepsilon),\]
of having leads polarized with respect to the band degree of freedom
$\tilde{r}$, might be fulfilled.

\section{\label{sec:The-matrix-elements}The matrix elements of the electron
operators}

In this appendix we calculate the expressions for the matrix elements
$\left\langle \vec{N},\vec{m}\right|\psi_{\tilde{r}\sigma F}(x)\left|\vec{N}',\vec{m}'\right\rangle $
using the bosonization identity (\ref{eq:bosident}) of the electron
operators $\psi_{\tilde{r}\sigma F}(x),$ namely\begin{equation}
\psi_{\tilde{r}\sigma F}(x)=\frac{\eta_{\tilde{r}\sigma}K_{\tilde{r}F\mathcal{N}_{\tilde{r}\sigma}}}{\sqrt{1-e^{-\alpha\frac{\pi}{L}}}}e^{i\phi_{\tilde{r}\sigma F}^{\dagger}(x)+i\phi_{\tilde{r}\sigma F}(x)}.\label{eq:bosident2}\end{equation}
As a reminder, $\eta_{r\sigma}$ is the Klein factor reducing the
electron number by one, the operator $K_{\tilde{r}F\mathcal{N}_{\tilde{r}\sigma}}$
is given by \begin{equation}
K_{\tilde{r}F\mathcal{N}_{r\sigma}}(x):=\frac{1}{\sqrt{2L}}e^{i\frac{\pi}{L}\textrm{sgn}F(\textrm{sgn}\tilde{r}\cdot\mathcal{N}_{\tilde{r}\sigma}+\delta)x}\label{eq:K-operator}\end{equation}
and yields a phase factor depending on the filling of the band $(\tilde{r}\sigma).$
The bosonic fields $\phi_{\tilde{r}F\sigma}(x)$ are given in terms
of the bosonic annihilation operators $b_{\sigma q}$,
\begin{equation}
i\phi_{\tilde{r}\sigma F}(x)=\sum_{q>0}\frac{e^{i\textrm{sgn}(F\tilde{r})qx-\alpha q/2}}{\sqrt{n_{q}}}b_{\sigma\textrm{sgn}(\tilde{r})q}.\label{eq:phifields}\end{equation}
The action of $\psi_{rF\sigma}(x)$ on the states $\left|\vec{N},\vec{m}\right\rangle $
is conveniently determined by rewriting the operators $b_{\sigma\textrm{sgn}(r)q}$
in terms of the operators $a_{j\delta q}$ and $a_{j\delta q}^{\dagger}$which
diagonalize the SWNT Hamiltonian $H_{\odot}$. As we know from the
Bogoliubov transformation, the $b$ operators depend linearly on the
$a$ operators and hence we can write in general\begin{multline}
i\phi_{\tilde{r}\sigma F}^{\dagger}(x)+i\phi_{\tilde{r}\sigma F}(x)=\\
=\sum_{q>0}e^{-\alpha q/2}\sum_{j\delta}\left(\lambda_{\tilde{r}\sigma F}^{j\delta q}(x)a_{j\delta q}+\lambda_{\tilde{r}\sigma F}^{\oplus j\delta q}(x)a_{j\delta q}^{\dagger}\right),\label{eq:iphiplusiphidagger}\end{multline}
where $\lambda_{\tilde{r}\sigma F}^{\oplus j\delta q}(x)=-\left(\lambda_{\tilde{r}\sigma F}^{j\delta q}(x)\right)^{*},$
since $i\phi_{\tilde{r}\sigma F}^{\dagger}(x)+i\phi_{\tilde{r}\sigma F}(x)$
is anti-hermitian. The actual values of the $\lambda$s are calculated
below in this appendix. Plugging the bosonization identity (\ref{eq:bosident2})
together with (\ref{eq:iphiplusiphidagger}) into $\left\langle \vec{N},\vec{m}\right|\psi_{\tilde{r}\sigma F}(x)\left|\vec{N}',\vec{m}'\right\rangle $,
yields\begin{multline*}
\left\langle \vec{N},\vec{m}\right|\psi_{\tilde{r}\sigma F}(x)\left|\vec{N}',\vec{m}'\right\rangle =\\
\delta_{\vec{N}+\vec{e}_{\tilde{r}\sigma},\vec{N}'}\frac{(-1)^{\sum_{j=1}^{\tilde{r}\sigma-1}N_{j}}}{\sqrt{1-e^{-\alpha\frac{\pi}{L}}}}K_{\tilde{r}F(\vec{N}')_{\tilde{r}\sigma}}(x)\times\\
\left\langle \vec{m}\right|e^{\sum_{q>0}e^{-\alpha q/2}\sum_{j\delta}\left(\lambda_{\tilde{r}\sigma F}^{j\delta q}(x)a_{j\delta q}+\lambda_{\tilde{r}\sigma F}^{\oplus j\delta q}(x)a_{j\delta q}^{\dagger}\right)}\left|\vec{m}'\right\rangle .\end{multline*}
Remember that the factor $(-1)^{\sum_{j=1}^{\tilde{r}\sigma-1}N_{j}}$
stems from the Klein factor. The term \begin{equation}
\left\langle \vec{m}\right|e^{\sum_{q>0}e^{-\alpha q/2}\sum_{j\delta}\left(\lambda_{\tilde{r}\sigma F}^{j\delta q}(x)a_{j\delta q}+\lambda_{\tilde{r}\sigma F}^{\oplus j\delta q}(x)a_{j\delta q}^{\dagger}\right)}\left|\vec{m}'\right\rangle \label{eq:bosonicpart}\end{equation}
 does not depend on the fermionic configuration and therefore we have
dropped the $\vec{N}$ index in $\left|\vec{N},\vec{m}\right\rangle $.
Now we exploit that in our finite size SWNTs there will always be
only a finite number of bosonic excitations. Hence, by using the Baker-Hausdorff
formula, $e^{A+B}=e^{A}e^{B}e^{-[A,B]/2}$, we commute all annihilation
operators $a_{j\sigma q}$ to the right and all creation operators
$a_{j\delta q}^{\dagger}$ to the left in (\ref{eq:bosonicpart}):\begin{multline}
\left\langle \vec{m}\right|e^{\sum_{q>0}e^{-\alpha q/2}\sum_{j\delta}\left(\lambda_{\tilde{r}\sigma F}^{j\delta q}(x)a_{j\delta q}+\lambda_{\tilde{r}\sigma F}^{\oplus j\delta q}(x)a_{j\delta q}^{\dagger}\right)}\left|\vec{m}'\right\rangle =\\
e^{-\frac{1}{2}\sum_{q>0}e^{-\alpha q}\sum_{j\delta}\left|\lambda_{\tilde{r}\sigma F}^{j\delta q}(x)\right|^{2}}\times\label{eq:separated}\\
\prod_{q>0}\prod_{j\delta}\left\langle m_{j\delta q}\right|e^{\lambda_{\tilde{r}\sigma F}^{\oplus j\delta q}(x)a_{j\delta q}^{\dagger}}e^{\lambda_{\tilde{r}\sigma F}^{j\delta q}(x)a_{j\delta q}}\left|m'_{j\delta q}\right\rangle ,\end{multline}
with $\left|m_{j\delta\kappa}\right\rangle =\left(m_{j\delta\kappa}!\right)^{-1/2}\left(a_{j\delta\kappa}^{\dagger}\right)^{m_{j\delta\kappa}}\left|0\right\rangle .$
Expanding the exponentials $e^{\lambda_{\tilde{r}\sigma F}^{j\delta q}(x)a_{j\delta q}}$
in (\ref{eq:separated}), all terms which are of higher order than
$m'_{j\delta q}$ will vanish. Analogously, all terms in the expansion
of $e^{\lambda_{\tilde{r}\sigma F}^{\oplus j\delta q}(x)a_{j\delta q}^{\dagger}}$
of higher order than $m_{j\delta q}$ can be ignored. Hence we get
\begin{multline}
\left\langle m\right|e^{\lambda^{\oplus} a^{\dagger}}e^{\lambda a}\left|m'\right\rangle =:F(\lambda,m,m')=\\
\left\langle m\left|\sum_{i=0}^{m}\frac{\left(\lambda^{\oplus}a^{\dagger}\right)^{i}}{i!}\sum_{j=0}^{m'}\frac{\left(\lambda a\right)^{j}}{j!}\right|m'\right\rangle .\label{eq:expandedmatrixelement}\end{multline}
Here, in favor of readability all indices have been dropped. In (\ref{eq:expandedmatrixelement})
only the terms with $m-i=m'-j$ survive. For $m'>m$ we conveniently
express $j$ in terms of $i$ and get:\begin{equation}
F(\lambda,m,m')=\lambda^{m'-m}\sqrt{\frac{m!}{m'!}}\sum_{i=0}^{m}\frac{\left(\lambda^{\oplus}\lambda\right)^{i}}{i!(i+m'-m)!}\frac{m'!}{(m-i)!}\label{eq:matrixelementmlessmpr}\end{equation}
and for $m'<m$ we write $i$ in terms of $j$ and get\begin{multline}
F(\lambda,m,m')=\\
\underbrace{\left(\lambda^{\oplus}\right)^{m-m'}}_{\left(-\lambda^{*}\right)^{m-m'}}\sqrt{\frac{m'!}{m!}}\sum_{j=0}^{m'}\frac{\left(\lambda^{\oplus}\lambda\right)^{j}}{j!(j+m-m')!}\frac{m!}{(m'-j)!}.\label{eq:matrixelementmprlessm}\end{multline}
Defining $m_{\mathrm{max}/\mathrm{min}}:=\max\textrm{/}\min(m,m')$, we can summarize
(\ref{eq:matrixelementmlessmpr}) and (\ref{eq:matrixelementmprlessm})
as \begin{multline}
F(\lambda,m,m')=\\
=\left(\Theta(m'-m)\lambda^{m'-m}+\Theta(m-m')\left(-\lambda^{*}\right)^{m-m'}\right)\\
\times\sqrt{\frac{m_{\mathrm{min}}!}{m_{\mathrm{max}}!}}\sum_{i=0}^{m_{\mathrm{min}}}\frac{\left(-\left|\lambda\right|^{2}\right)^{i}}{i!(i+m_{\mathrm{max}}-m_{\mathrm{min}})!}\frac{m_{\mathrm{max}}!}{(m_{\mathrm{min}}-i)!},\label{eq:Fexpl1}\end{multline}
or in terms of the Laguerre polynomials \cite{gradshteyn2000},\begin{multline}
F(\lambda,m,m')=\sqrt{\frac{m_{\mathrm{min}}!}{m_{\mathrm{max}}!}}L_{m_{\mathrm{min}}}^{m_{\mathrm{max}}-m_{\mathrm{min}}}(\left|\lambda\right|^{2})\times\\
\left(\Theta(m'-m)\lambda^{m'-m}+\Theta(m-m')\left(-\lambda^{*}\right)^{m-m'}\right).\label{eq:Fexpl2}\end{multline}
Similar expressions for $F(\lambda,m,m')$ were found by Kim et al.
\cite{Kim2003} when investigating charge plasmons in a spinless Luttinger
liquid quantum dot. In the end we obtain for the matrix elements $\left(\psi_{\tilde{r}\sigma F}(x)\right)_{nm}$
the following expression:\begin{multline*}
\left\langle \vec{N},\vec{m}\right|\psi_{\tilde{r}\sigma F}(x)\left|\vec{N}',\vec{m}'\right\rangle =\\
\delta_{\vec{N}+\vec{e}_{\tilde{r}\sigma}\vec{N}'}(-1)^{\sum_{j=1}^{\tilde{r}\sigma-1}N_{j}}K_{\tilde{r}F(\vec{N}')_{\tilde{r}\sigma}}(x)A(x)\times\\
\prod_{q>0}\prod_{j\delta}F(\lambda_{\tilde{r}\sigma F}^{j\delta q}(x),m_{j\delta q},m'_{j\delta q}),\end{multline*}
where we have defined \[
A(x):=\frac{e^{-\frac{1}{2}\sum_{q>0}e^{-\alpha q}\sum_{j\delta}\left|\lambda_{\tilde{r}\sigma F}^{j\delta q}(x)\right|^{2}}}{\sqrt{1-e^{-\alpha\frac{\pi}{L}}}}.\]

\section{\label{sec:The-parameters-}The parameters $\lambda$ }

We still have to determine the values of the parameters $\lambda_{r\sigma F}^{j\delta q}(x)$
for our case. Using equation (\ref{eq:phifields}) we can express
the sum $i\phi_{r\sigma F}^{\dagger}(x)+i\phi_{r\sigma F}(x)$ in
terms of the bosonic operators $b_{\sigma q},$\begin{multline}
i\phi_{r\sigma F}^{\dagger}(x)+i\phi_{r\sigma F}(x)=\sum_{q>0}\frac{e^{-\alpha q/2}}{\sqrt{n_{q}}}\times\\
\left(-e^{-i\textrm{sgn}(Fr)qx}b_{\sigma\textrm{sgn}(r)q}^{\dagger}+e^{i\textrm{sgn}(Fr)qx}b_{\sigma\textrm{sgn}(r)q}\right),\end{multline}
 which in turn are related to the operators $a_{j\delta q}$ and $a_{j\delta q}^{\dagger}$
via equation (\ref{eq:bintermsofas}). Hence \begin{multline}
i\phi_{r\sigma F}^{\dagger}(x)+i\phi_{r\sigma F}(x)=\\
\sum_{j\delta}\Lambda_{\sigma\tilde{r}}^{j\delta}\sum_{q>0}\frac{e^{-\alpha q/2}}{\sqrt{n_{q}}}\left[e^{i\textrm{sgn}(Fr)qx}\left(C_{j\delta q}a_{j\delta q}^{\dagger}+S_{j\delta q}a_{j\delta q}\right)\right.\\
\left.-e^{-i\textrm{sgn}(Fr)qx}\left(S_{j\delta q}a_{j\delta q}^{\dagger}+C_{j\delta q}a_{j\delta q}\right)\right]\label{eq:iphifieldsintermsofa}\end{multline}
By comparing (\ref{eq:iphifieldsintermsofa}) to (\ref{eq:iphiplusiphidagger})
the values of the $\lambda$s can be read off:\begin{multline*}
\lambda_{\tilde{r}\sigma F}^{j\delta q}(x)=\\
\frac{1}{\sqrt{n_{q}}}\Lambda_{\tilde{r}\sigma}^{j\delta}\left(-e^{-i\textrm{sgn}(Fr)qx}C_{j\delta q}+e^{i\textrm{sgn}(Fr)qx}S_{j\delta q}\right).\end{multline*}
With the coefficients $C_{j\delta q},\, S_{j\delta q}$ and $\Lambda_{\sigma\tilde{r}}^{j\delta}$
from (\ref{eq:Lambdajdelta_sigmar}), (\ref{eq:Sneutral}) and (\ref{eq:Scplus})
we get for $j\delta=c-,s+,s-,$
\begin{equation}
\lambda_{\tilde{r}\sigma F}^{j \delta q}(x) =\frac{e^{i\mathrm{sgn}(F\tilde{r})qx}}{\sqrt{n_{q}}}\Lambda_{\tilde{r}\sigma}^{j\delta}\label{eq:lambdaneutral2}\end{equation}
 and for the $c+$ mode we have\begin{multline}
\lambda_{\tilde{r}\sigma}^{c+q}(x)=\\
\frac{1}{2\sqrt{n_{q}}}\left(\sqrt{\frac{\varepsilon_{cq}}{\varepsilon_{0q}}}\cos(qx)+i\sqrt{\frac{\varepsilon_{0q}}{\varepsilon_{cq}}}\mathrm{sgn}(F\tilde{r})\sin(qx)\right).\label{eq:lambdacharged2}\end{multline}

\end{document}